\documentclass{article}
\usepackage[a4paper, total={6in, 10in}]{geometry}
\usepackage[utf8]{inputenc}
\usepackage{amsmath}
\usepackage{hyperref}
\usepackage{graphicx}
\usepackage{lipsum}
\usepackage{xspace,graphics,verbatim,amsmath,amsfonts}
\makeatletter
\newcommand*{\rom}[1]{\expandafter\@slowromancap\romannumeral #1@}
\makeatother

\hypersetup{
    colorlinks=true, 
    linkcolor=red,   
    citecolor=blue,  
    filecolor=blue,  
    urlcolor=blue    
}

\newcommand{\ben}{\begin{equation}}
\newcommand{\een}{\end{equation}}
\newcommand{\bean}{\begin{eqnarray}}
\newcommand{\eean}{\end{eqnarray}}

\newcommand{\bea}{\begin{eqnarray*}}
\newcommand{\eea}{\end{eqnarray*}}

\usepackage{lineno}
\usepackage{hyperref}
\usepackage{upref}
\usepackage{enumerate}

\usepackage{appendix}
\usepackage{graphicx}
\usepackage{caption}
\usepackage{subcaption}
\usepackage{epsfig}
\usepackage{color}
\DeclareGraphicsExtensions{.pdf,.png,.jpg}

\usepackage{tikz}
\usepackage{xspace}
\usepackage{amssymb,amsmath,latexsym,wasysym,mathrsfs,amsfonts,amsthm,bbm}
\usepackage{yfonts,amscd,amsxtra}
\usepackage{amstext,verbatim}
\usepackage{bbm}

\usepackage{multicol}
\usepackage{float}
\usepackage{array}
\usepackage{multirow}       
\usepackage{siunitx}
\restylefloat{table}

\newcommand{\bes}{\begin{equation*}}
\newcommand{\ees}{\end{equation*}}
\newcommand{\beq}{\begin{equation}}
\newcommand{\eeq}{\end{equation}}
\newcommand{\barr}{\begin{array}}
\newcommand{\earr}{\end{array}}

\usepackage[makeroom]{cancel}
\theoremstyle{plain}

\theoremstyle{definition}

\theoremstyle{remark}

\usepackage{pgf,tikz}
\usepackage{mathrsfs}
\usetikzlibrary{arrows}
\usetikzlibrary{matrix}
\usepackage{dsfont}

\usepackage{url}

\usepackage{pgfplots}
\usepackage{tikz}
\usepackage{pgf}
\usepackage{mathrsfs}
\usetikzlibrary{calc}
\usetikzlibrary{arrows}
\usetikzlibrary{matrix}
\pgfplotsset{compat=1.16}
\usepackage{float}
\usepackage{multirow} 
\restylefloat{table}
\usepackage{graphicx}

\usepackage{multicol}

\usepackage{graphicx}
\begin{document}
\vspace*{0.2in}

\begin{flushleft}
{\Large
\textbf\newline{The effect of environmental factors in biofilm‑phage interactions in Agent Based Model} 
}
\newline
\\
Blessing O. Emerenini \textsuperscript{1*},
Edward Beck\textsuperscript{2},
Kathryn Cantrel\textsuperscript{3},
Vedat Kurtay\textsuperscript{1},

\bigskip
\textbf{1} School of Mathematics and Statistics, College of Science, Rochester Institute of Technology, Rochester, NY, USA
\\
\textbf{2} University of Florida, Gainesville, Florida, USA
\\
\textbf{3} Loyola University Chicago, Illinois, Chicago, USA
\\
\bigskip






* boesma@rit.edu

\end{flushleft}
\section*{Abstract}
As antibiotic resistance continues to pose a significant threat to public health, alternative treatments are urgently needed. Phage therapy, which utilizes bacteriophages to specifically target bacterial pathogens, has emerged as a promising solution. Given that bacteria often exist in biofilms -complex micro-communities that complicate treatment strategies, there is a clear need for models that account for spatial dynamics. This study aims to employ mathematical and statistical methodologies to identify optimal treatment strategies involving phage-antibiotic combinations. We developed an agent-based model to analyze how environmental factors (e.g. temperature, pH, and resource availability) influence bacteria-phage interactions during therapy, focusing on both healthy and immunocompromised patients. Utilizing \textit{Escherichia coli} as a case study, we observed that bacterial cells exhibit mutations that enhance their adaptability to varying environmental conditions and treatment approaches. Our findings suggest that the effectiveness of therapies targeting pathogenic and mutated bacterial cells can be significantly improved through strategic control of application timing and dosing. Additionally, we investigated the impact of biofilm structure on the efficacy of phage therapy, underscoring its importance in developing targeted treatment strategies.

\vspace{.2in}

\noindent \textbf{Key words:} Phages, Biofilm, Antibiotics, Agent based model, Corrosion, NetLogo

\section{Introduction}
Bacterial infections are a significant cause of various diseases, resulting in millions of global human deaths each year \cite{Ikuta2019}. To combat these infections, the development and prescription of new antibiotics have been prevalent over the past century. However, bacteria have evolved to become resistant to these drugs through genetic mutation and natural selection, leading to a decline in the effectiveness of many standard-of-care antibiotics \cite{Dutescu2021, Kim2018}. Antibiotic resistance has been recognized as a global public health crisis by leaders in the field, posing a threat to our ability to combat diseases effectively \cite{Lim2010, WHO2014}. It is estimated that antimicrobial resistance caused around 4.95 million deaths globally in 2019, with approximately 1.27 million of these deaths attributed to antibacterial resistance \cite{murray2021}. In the United States alone, antibiotic resistance leads to more than 2.8 million infections and over 35,000 deaths annually \cite{CDC2019}. Additionally, the use of antibiotics can harm both infection-causing and beneficial bacteria, necessitating the exploration of alternative methods to combat bacterial infections \cite{Patangia2022, Muhammad2020, Melander2018}.

One such alternative treatment involves bacteriophages commonly referred to as 
phages, which are viruses that infect bacteria and utilize their metabolism to produce more phage copies, ultimately leading to the death of the infected bacterial cell \cite{Guang-Han2016}. Phage therapy, which utilizes these phages to target bacterial infections in humans, has garnered attention as a potential solution due to its specificity and ability to self-multiply at the site of infection, minimizing collateral damage \cite{Pires2021}. Moreover, phages are abundant in nature, making them readily available for medical use. With the increasing concerns about antibiotic resistance, phage therapy has gained renewed interest as a potential alternative or supplement to antibiotic treatments.

Bacteria most commonly occur in biofilms, biological structures which attach to a surface and offer protection from antimicrobial agents or harsh environments \cite{Pires2021}. When a biofilm continues to gain more area, it becomes more difficult for an antimicrobial to eradicate it from an infected organism. Within the biofilms, the typically pathogenic bacterial cells become attached, or sessile, to their surface and create extracellular polymeric substances, also referred to as EPSs, which shield the pathogens from antibiotics or other antimicrobials. While the study of bacterial infections has a history dating back thousands of years, biofilms have only been explored since the 17th century \cite{chandki2011}.

The human immune system, particularly innate immunity, plays a crucial role in eradicating bacterial infections and biofilms \cite{Leung2017}. Innate immunity provides a faster and more general response to pathogens compared to adaptive immunity, which is slower and more targeted. Understanding the interactions between antibiotics, phages, immunity, and pathogenic bacteria is essential for comparing treatment strategies effectively \cite{Leung2017}.

In this study, we aimed to create a model that simulates the behaviors of antibiotics, phages, and bacteria while using biologically motivated values for their movements and functions. The objective was to explore various environments and their effects on pathogenic growth rates to determine effective treatment strategies. Although both innate and adaptive immunity are vital in eliminating pathogens, we focused on short-term infection dynamics in this study and did not consider the impact of the adaptive immune system.

Various modeling approaches have been utilized to explore interactions between antibiotics, phages, and bacteria. For the current study, both qualitative and quantitative results are essential. While ordinary differential equation (ODE) models are commonly used in computational biology, they are limited to single independent variables, making them unsuitable for considering multiple independent variables in multiple species \cite{emerenini2021}. Partial differential equation (PDE) models, while useful for physical problems, lack the clarity needed for our model's physical parameters. Consequently, we opted for Agent-Based Modeling (ABM) to simulate interactions between different species with distinct properties, accounting for spatial effects and individual-level changes within the population \cite{Hellweger2016, di_martino2018}. ABM is a stochastic modeling system that employs a bottom-up approach, where population-level behaviors emerge from individual entities rather than the other way around \cite{di_martino2018}. ABM has been successfully applied to study bacterial and biofilm growth dynamics, as well as the effects of antibiotics and phage treatments \cite{Murphy2011, Heilmann2012, Eriksen2018, Koonin2015}. Incorporating immunity and other human effects, our ABM model extends current research on phage therapy. Despite the limitations of Agent-Based Models, such as the need to interpret data for numerous agents at each time step, the impact is negligible for the scope of this study, which solely aims to observe short-term behaviors within a confined space.

Throughout this article, we will present the parameters, numerical formulations, and simulations used in our model. We will discuss the assumptions made and then present the data from our simulations involving antibiotics, phages, and their combined treatments in both free-floating and biofilm settings. We will also explore the impact of environmental conditions on each treatment. Finally, we will interpret the results in the context of biology and medicine, determine the most effective treatment strategy based on performance metrics, and compare the outcomes of our three simulations.

\section{Methods and Model Formulation}

\subsection{Model Basics and Parameters}


We developed an agent-based model in the NetLogo program that accounts for the numerous factors at play in the growth of infectious bacteria in human or \emph{in vitro} settings. NetLogo is a programming language and environment developed in 1999 and based on the Logo language \cite{NetLogo}. It features an interface that includes a 2- or 3- dimensional display of the ``world" in which the simulation plays out, optional plots and monitors for outputs, and buttons and other interactive elements to build an easy-to-operate front-end. The model updates in discrete timesteps called ticks. The world consists of a grid of square regions, called patches, with individual properties. Agents are created and given individual properties, move in specified patterns, and interact with each other and their environment according to rules specified in sections of code called procedures. These features make NetLogo ideal for modeling the dynamics of bacterial growth and elimination, where spatial effects and population heterogeneity are essential.


Four types of agents were modeled, named in the model as: pathogens (infectious bacteria), phages, antibiotics, and neutrophils (human innate immune cells). The environment was laid out in a grid of regions, called patches, which could contain a number of nutrients (sugar) on which the pathogens feed. Patches correlate to  $2.0 \mu m$ by $2.0 \mu m$ squares, chosen so that an \emph{E. coli} cell, the archetypical bacterial pathogen shaped as a rough cylinder typically of maximum length $2.0 \mu m$, should occupy approximately one whole patch \cite{Riley1999}. A square environment $49$ patches ($98 \mu m$) long on a side was chosen to fully explore the complex spatial dynamics expected from biofilm-phage interactions. Ticks were chosen to represent half of one minute to be on the order of magnitude of most bacterial generation times while being sufficiently small to replicate the maximum growth rates observed in experiments. For every tick, agents have the potential to move, reproduce, consume nutrients, and/or interact with other agents, according to the real behavior of the biological entities. Formulation of the model was in part informed by an ABM text for microbial systems using NetLogo \cite{Yust2020}.


Each type of agent is assigned realistic values for a set of biologically relevant parameters upon creation. For pathogens, the values of some parameters are allowed to change from one generation to the next with a specified probability, equivalent to the rate of genetic mutation. Including individual variation and mutation allows the modeling of infected populations, adaptation to changing environments, and evolution of bacterial resistance to antibiotic and phage treatments. This is a major advantage of ABM over methods that assume uniform properties across a population.

All agents are given the properties of health, available energy, energy cost of movement, energy requirement for division, and minimum and maximum temperature and pH values for growth (see Table 1 for assigned base values). Bacterial pathogens are characterized by additional properties for energy cost of survival (due to ongoing metabolic processes), antibiotic resistance, phage immunity acquisition, phage infection status, rate of biofilm development, and cellular capacity to produce phages. Properties of pathogens that may mutate include the energy cost of movement, energy requirement for division, minimum and maximum temperature and pH values for growth, antibiotic resistance, and capacity to produce phages. Phages have additional properties for burst size, latency time of infection, decay rate, adsorption rate, and rate of degradation of the EPS matrix. Antibiotics are characterized by a rate of damage to adjacent pathogens and a decay rate and may be globally specified as broad-spectrum or narrow-spectrum. Neutrophils have additional properties for rate of phagocytosis (uptake and killing of pathogens) and lifespan. Properties related to health, energy, and viable environmental conditions are irrelevant to all non-pathogen agents and only included to conveniently use the same procedures for all agents.


A broad set of parameters is necessary to completely define our model. A full list is presented in Table 1, along with the values chosen in the baseline numerical simulation. These parameters include tick duration, environmental properties of temperature and pH, generation rate and maximum count per patch for nutrients, pathogen mutation rate, populations/doses of agents created, rates and population counts defining the appearance and disappearance of antibiotics and neutrophils, and base values for the individual properties of the different agents listed previously.

\subsection{Model Assumptions}


The human small intestine was chosen as the environment for the \emph{in vivo} simulations since it is often the site of bacterial infection as well as much absorption of drugs and nutrients \cite{Koziolek2014}. Temperature, pH, and nutrient availability were selected as the most important environmental factors in bacterial growth and were therefore included in our model \cite{Yust2020, Ross2003, Kumar2013}. \emph{Escherichia coli (E. coli)} was chosen as the bacterial pathogen for our model because of its frequent cause of infection in the human intestine and due to the wealth of literature surrounding its properties and treatment \cite{Ross2003, Kumar2013, Yust2020, Riley1999, Drake1998}. Furthermore, phage therapy has been developed for treatment of \emph{E. coli} \cite{Ferry2021}. Specifically, bacteriophage $\lambda$ targets \emph{E. coli} and was therefore chosen as the type of phage in our model \cite{brown2022}. Here we only simulate general properties of antibiotics, but the antibiotic agents in our model could represent the broad-spectrum antibiotic Rifaximin, which is often used to treat \emph{E. coli}. Immune cells in our model represent neutrophils, which are a major component of the human innate immune response and play a role in the success of phage therapy \cite{Mayadas2014, Leung2017}.

We attempt to model the short-term dynamics of bacteria, phages, antibiotics, and innate immunity in a small two-dimensional region. 

\subsubsection*{\textit{Assumption on agents:}}
We therefore assume that the activities of the agents are restricted to a bounded plane and that agents cannot leave the space unless they die or decay. This is equivalent to assuming that a larger region of infection exists outside of the simulated region and that agents enter from the simulated region and exit to the larger region at equal rates.

The spatial restrictions of our model result in a much reduced simulated population and infection area compared with real bacterial infections and treatments. A reduced population requires a higher rate at which mutable properties may change than would be realistic in order to observe the effects of evolution. Bacterial pathogens are prevented from overlapping, which causes increasing density to limit their growth since space is restricted. Phage growth is also limited through a smaller burst size than the realistic value in order to restrict the sizes of the modeled populations and avoid excessive computation. Values for the numbers of antibiotics and phages used in treatments are arbitrary, since it is infeasible to relate those quantities to real doses; however, we consider their relative sizes to be informative for assessing the comparative effectiveness of different combinations of phage and antibiotic doses. Assumptions specific to the individual agents are listed as follows:
\begin{itemize}
    \item \textbf{Pathogens} may not overlap with each other as they move and divide.  Pathogens increase the hydrogen ion concentration of the world at a rate proportional to their population. This captures the effect of mixed-acid fermentation performed by \emph{E. coli} on the acidity of their environment \cite{Yust2020}. Bacterial infections are known to cause local changes in pH within the human body \cite{Bullock2020}.
    \item \textbf{Antibiotics} are subject to the effects of dilution and degradation, represented by a constant half-life \cite{Wishart2006, Murphy2011}.
    \item The \textbf{adaptive immune response} plays a negligible role since the simulation accounts only for the short-term (12-hour) dynamics of infection and treatment \cite{Leung2017}. The innate immune response consists only of neutrophils, which enter the simulated region of the intestine from elsewhere in the body as pathogens grow. A neutrophil may be at any point in its lifespan when it appears. The innate immune response saturates beyond a certain level of activity due the limited number of available neutrophils \cite{Leung2017}. In immunocompromised settings, the innate immune response has a negligible effect and no neutrophils enter the small simulated region.
    \item \textbf{Phage} population present within the computational domain is assumed to be limited, since only a finite number of real phages may occupy a restricted region of space. All phages initially created are genetically identical. Every phage newly created through lysis has a chance of mutating, resulting in a new genome \cite{Childs2012}. When a phage attempts to infect a pathogen, the pathogen may acquire immunity to all phages of the same genome. This captures the effect of the CRISPR-Cas adaptive immune system present in \emph{E. coli}, by which bacterial cells develop the ability to target foreign viral DNA \cite{Koonin2015, Childs2012}. 
\end{itemize}
Division, mutation, adsorption, phage and antibiotic decay, neutrophil lifespan, phagocytosis, bacterial phage immunity acquisition, initial agent position, and nutrient spatial distribution are all probabilistic processes. This serves to capture the stochasticity inherent at the individual level.
\subsubsection*{\textit{Assumption on environmental factors: }} The pH is automatically regulated by the human body towards its typical value in the small intestine \cite{pHRegulation}. This occurs with a strength proportional to the difference between the target hydrogen ion concentration and the current concentration.

\subsubsection*{\textit{Biofilm Model Assumptions}}
In our model development, we have also considered the effect of bacteria-phage interaction on biofilm development and structures. The biofilm is allowed to develop along the bottom border of the world and grow outward to a given extent during or prior to treatment, as modeled in \cite{Emerenini2015}. This bottom border represents an inner surface of the small intestine on which infectious bacteria could adhere and grow. Here we list additional assumptions made for this model:

\begin{enumerate}


    \item A planktonic pathogen may become sessile (part of a biofilm) if it is located near the bottom border of the world or an existing sessile pathogen. On the other hand, a sessile pathogen may become planktonic if all surrounding pathogens are planktonic, i.e., it is no longer within a biofilm.
    \item Sessile pathogens may not move but may continue to divide, and their division is less restricted by density than that of planktonic pathogens. Division deep within the biofilm is still highly restricted by density, which may be justified by the dormant (low-metabolic activity) state that characterizes older biofilm cells \cite{Pires2017, Pires2021}.

    \item Sessile pathogens have lower metabolic activity and thus a lower energy expenditure for metabolic processes and, if infected, a lower capacity for replicating phages \cite{Pires2017, Pires2021, Heilmann2012}.

    \item Phages located within a biofilm or near sessile pathogens have a higher decay rate, lower adsorption rate, and lower mobility, proportional to the number of nearby sessile pathogens. This accounts for the defensive properties of biofilms and the EPS matrix against attacking phages \cite{Pires2021, Heilmann2012}. A phage located near an outer edge of a biofilm may convert a nearby sessile pathogen to a planktonic pathogen. This represents phages' ability to produce depolymerases that degrade EPS and thus break down biofilm structures \cite{Pires2021}.

    \item Planktonic pathogens are diluted out of the environment at a deterministic rate proportional to their population and approximately equal to their growth rate. This is done in order to model a constant bacterial density within the environment surrounding the developing biofilm.

\item A tolerance is put in place to ensure that the biofilm does not grow unbounded and has a finite speed of propagation \cite{Emerenini2015,emerenini2022}


    \item Adhesion and EPS degradation are probabilistic processes. This serves to capture the stochasticity inherent at the individual level.
    
\end{enumerate}

Note that due to the additional dilution rate in the biofilm model, the pathogen populations cannot be accurately compared between the original and biofilm models. However, the relative effects of different treatment strategies within each model remain significant.

\subsection{Formulations}


\subsubsection*{Bacterial Relative Growth Rate Formula}

We use the function for relative growth rate of \emph{E. coli} depending on environmental factors developed in \cite{Ross2003}, which describes fitting a model to experimental data. We remove all factors not dependent on temperature or pH, resulting in an equation for relative growth rate as a function of those two variables. We restrict the domain of the function to the ranges of temperature and pH between the reported minimum and maximum values of each for possible growth. We then scale the function by a constant so that the relative growth rate at the optimal values of temperature and pH results in the theoretical 20-minute doubling time for \emph{E. coli} under ideal conditions (see Appendix for full derivation) \cite{gibson2018}.

We adopt the relative growth rate in \cite{Ross2003}, written as follows: 

 \begin{equation} \label{RossRGR}
\hspace{-0.8in} f(T, pH) = {\left( \eta_1(T - T_{min}) \left( 1 - \exp{ \left( \eta_2 (T - T_{max}) \right) } \right) \right) }^{2} \left(1 - {10}^{pH_{min} - pH}\right) \left(1 - {10}^{pH - pH_{max}}\right)
\end{equation}
where $\eta_1=0.043885$ and $\eta_2 = 0.2636$; $T_{min}$ is the minimum temperature while $T_{max}$ is the maximum temperature. The parameter $pH_{min}$ is the minimum pH value while $pH_{max}$ is the maximum pH.

We use the following equation to convert from a deterministic formula $f(\Phi)$ for relative growth rate per hour, for some set of environmental parameters $\Phi=\{T,pH,...\}$, such as \eqref{RossRGR}, to the equivalent proportion $g(\Phi)$ of the population that divides each tick, where $q$ is the conversion factor from ticks to hours (see Appendix for derivation):

\begin{equation} \label{portionDivide}
g(\Phi) = e^{\frac{1}{q} \cdot f(\Phi)} - 1
\end{equation}

If every individual pathogen has a probability of $g(\Phi)$ of dividing each tick, then the expected proportion of the population that divides each tick is simply $g(\Phi)$, so we use \eqref{portionDivide} setting $f(\Phi)$ equal to the function in \eqref{RossRGR} as the equivalent probability of dividing per tick for each pathogen as follows (see Table 1 for value of $q$):

\begin{equation} \label{probDivide}
\hspace{-0.3in}  g(T, pH) = \exp{\left[ \left( \frac{1}{q} \right) {\left( \eta_1 (T - T_{min}) \left( 1 - \exp{ \left( \eta_2 (T - T_{max}) \right) } \right) \right) }^{2} \left(1 - {10}^{pH_{min} - pH}\right) \left(1 - {10}^{pH - pH_{max}}\right) \right]} - 1
\end{equation}

The function's maximum value, computed with parameters as outlined in Table 1, is $0.0175$, which is well below unity, thus making it appropriate for utilization as a probability measure. While there exists the possibility for bacteria to undergo mutations at extreme values of $T_{min}$, $T_{max}$, $pH_{min}$, and $pH_{max}$ resulting in a higher maximum value for the function $g$, such occurrences are exceedingly rare, and it is uncommon for this value to surpass $1$. In the context of the model, any value exceeding $1$ for this function is uniformly interpreted as a probability of $1$.

It is important to note that the function $g$ exhibits a consistent increase throughout its domain with diminishing values of $T_{min}$ and $pH_{min}$, and with elevated values of $T_{max}$ and $pH_{max}$. This characteristic not only delimits the range within which a pathogen may proliferate but also signifies a general adaptability to division under diverse environmental conditions, as represented by the parameters for minimum and maximum survivable temperature and pH.



\subsubsection*{Timescale Selection}

Since $g(\Phi)$ is the probability of dividing each tick, it must take on values less than or equal to 1. Using this goal and from \eqref{portionDivide}, we require that, for all $\Phi$,

\begin{equation} \label{qRestriction}
    f(\Phi) \le q \ln 2
\end{equation}

The maximum value of $f$ is $2.058$, so we require $q \ge 2.969$. However, a value of $q$ close to this number could also result in $g > 1$ due to mutations in survivable temperature and pH ranges, so a much larger value of $q$ (smaller time step) was desired. Half-minutes ($q=120)$ were therefore chosen as the real value for the time steps, or ticks, in the simulation; these are small enough intervals to nearly always avoid $g > 1$ and be on the order of magnitude of the generation time of \emph{E. coli} under ideal conditions, while avoiding an excessively small and therefore computationally intensive time step.

\subsubsection*{Decay Formula}

Some experimental decay rates adopted for our model were provided as hourly rates. To convert an hourly decay rate to an individual probability of decaying per tick, we used the following formula, where $\delta_t$ represents the decay rate in 1/hours and $\delta_\tau$ represents the equivalent decay rate in 1/ticks, equivalent to the probability of an individual agent decaying each tick (see Appendix for derivation):

\begin{equation} \label{ticklyRate}
\delta_\tau = 1 - {(1 - \delta_t)}^{1/q}
\end{equation}

A more detailed overview of the functioning of the various versions of the simulation is included in the Appendix.



\section{Numerical Simulation}
\subsection{Parameter Values}

Model parameter values were either selected from various experimental sources or assumed based on general biological principles (see Table 1) \cite{Bonilla2020, DePaepe2006, fallingborg1999, kannoly2023, pearson2012, Ross2003, Yust2020, Drake1998, Pires2017}. For our baseline results, temperature and pH were set to typical values for the human small intestine to simulate realistic conditions for growth of bacterial pathogens \cite{fallingborg1999, pearson2012, Koziolek2014}.
Base minimum and maximum values for the ranges of temperature and pH in which the pathogens may reproduce were pulled from a study on the effects of environmental factors on the growth of \emph{E. coli}, as was the environmental growth rate function used in the model \cite{Ross2003}. The temperature and pH of the human small intestine were taken from typical reported values, where the pH used is the average of the values at the entrance and exit of the small intestine \cite{pearson2012, Koziolek2014}.
A pathogen mutation rate of $0.25\%$ per genotype per generation was selected, using the documented rate of 0.25\% base pairs per genotype per generation for \emph{E. coli} \cite{Drake1998}. 

The quantification of parameters relevant to pathogen health, energy, nutrient abundance, and distribution was performed based on well-established scientific principles. These considerations included factors such as bacterial growth, energy expenditure for movement and metabolic processes, antibiotic-mediated bacterial elimination, and the formation and degradation of biofilms by bacteria and phages \cite{Yust2020, Patangia2022, Pires2021}. In the simulation, the nutrient generation rate was set to $1/8$ of the total number of patches, mimicking a continuous but stochastic nutrient supply. The base bacterial capacity to produce phages, denoted as $k_{max}$, was assumed to be $1$, implying that initially, all pathogens would release exactly one burst size of infecting phages upon infection and lysis. This assumption simplifies the representation of burst size from existing modeling literature \cite{kannoly2023}. Additionally, the rate of increase in hydrogen ion concentration, dependent on the pathogen population, was chosen as ${10}^{-9}$ per pathogen per tick, approximating a simplified function for hydrogen ion concentration change based on previous research\cite{Ratzke2018}.


Base parameter values for phages were taken from a characterization of bacteriophage $\lambda$, except for adsorption rate, which was increased so that bacteria-phage interactions could be observed in a reduced population within a reasonable time, and burst size, which was decreased in order to reduce simulated population sizes and save computation \cite{DePaepe2006}. Values for the dose period and half-life of antibiotics were taken from data on Rifaximin \cite{Rifaximin, DrugBank}. Neutrophil lifespan was reported in \cite{Bonilla2020}. Other parameter values for phages, antibiotics, and neutrophils (mobility, dose, activation threshold, etc.) were assumed from their qualitative behavior \cite{Pires2021, Murphy2011, Mayadas2014}. Parameters related to the formation of biofilms and their interaction with phages were also assumed from descriptions of those processes \cite{Pires2021, Pires2017, Heilmann2012, Schluter2015}.

 
\begin{table}[!b]
		\caption{\textbf{List of Parameter values and sources:} 
        \newline ** Parameter only relevant to bacterial pathogens; assigned values for other agents such that health, energy, and environmental factors have no effect (e.g. movement cost of 0 for non-pathogens).
        \newline *** Chosen to approximate simplified function from \cite{Ratzke2018} for hydrogen ion concentration change due to population and current pH.
        \newline $\dagger$ 1000 times greater than that reported in \cite{Childs2012} to observe immunity acquisition to phages in a reduced population.
        \newline $\ddagger$ Reduced from the value of 115 reported in \cite{DePaepe2006} in order to save computation.}
	\label{Table 1}
	\centering
	\begin{tabular}  {|p{2.3cm}|p{6.7cm}|p{1.8cm}|p{3cm}|p{1.5cm}|}
		\hline
		Symbol & Description & Value & Units & Source \\
		\hline
  		$q$ & ${\text{Tick Duration}}^{-1}$ & 120 & ${\text{hours}}^{-1}$ & selected (see SI) \\ [0.2ex]
            \hline
  		$T$ & Temperature of Environment & 39.6 & $^{\circ}$C & \cite{pearson2012} \\ [0.2ex]
		\hline
		$pH$ & pH of Environment & 6.7 & & \cite{fallingborg1999, Koziolek2014} \\ [0.2ex]
		\hline
  		$\tau_A$ & Antibiotic Dose Period & 960 (8) & ticks (hours) & \cite{StatPearls} \\ [0.3ex]
		\hline
    	$h_A$ & Antibiotic Half-Life & 6 & hours & \cite{Wishart2006} \\ [0.3ex]
		\hline
  		$DT$ & Bacterial Optimal Doubling Time & 20 & minutes & \cite{gibson2018} \\ [0.3ex]
		\hline
            $T_{GI}$ & Temperature of Human Small Intestine & 39.6 & $^{\circ}$C & \cite{pearson2012}\\ [0.2ex]
		\hline
            $pH_{GI}$ & pH of Human Small Intestine & 6.7 &  & \cite{Koziolek2014} \\ [0.2ex]
		\hline
		$T_{min}$, $T_{max}$ & Base Min/Max Survivable Temperature** & 4.14, 49.55 & $^{\circ}$C & \cite{Ross2003}  \\[0.2ex]
		\hline
		$pH_{min}$, $pH_{max}$ & Base Min/Max Survivable pH** & 3.909, 8.860 & & \cite{Ross2003} \\ [0.2ex]
		\hline
            $k_{max}$ & Base Bacterial Capacity to Produce Phages & 1 & & \cite{kannoly2023} \\ [0.3ex]
            \hline
            $\Delta_p$ & Acidity Increase Rate & ${10}^{-9}$ & pathogen/tick & \cite{Ratzke2018}*** \\ [0.3ex]
            \hline
            $r_{imm}$ & Immunity Acquisition Rate to Phages & ${10}^{-2}$ & count/interaction & \cite{Childs2012}$\dagger$ \\ [0.3ex]
            \hline
            $\beta$ & Burst Size & 5 & count & \cite{DePaepe2006}$\ddagger$ \\ [0.3ex]
            \hline
            $t_{L}$ & Latency Time & 84 (42) & ticks (minutes) & \cite{DePaepe2006} \\ [0.3ex]
            \hline
            $\delta_P$ & Phage Decay Rate & 0.072 & count/day & \cite{DePaepe2006} \\ [0.3ex]
            \hline
            $spec$ & Antibiotic Spectrum (0 or 1) & 0 (broad) & & \cite{Rifaximin} \\ [0.3ex]
            \hline
            $L_N$ & Maximum Neutrophil Lifespan & 2880 (1) & ticks (days) & \cite{Bonilla2020} \\ [0.3ex]
            \hline

	\end{tabular}\label{table1}
\end{table}

\begin{table*}[!h]
\begin{center}
\caption{\textbf{Default initial conditions, base parameter values and rates.} These parameters are based on our model assumptions. ``Base" refers to values assigned at agent creation and subject to change during the simulation.
 \newline ** Parameter only relevant to bacterial pathogens; assigned values for other agents such that health, energy, and environmental factors have no effect.}
\label{params}
\begin{tabular}{lccc}
\hline \hline
symbol & parameter & value  & units \\
\hline \hline
$B_0$ & Initial Bacterial Count & 100 & count\\ [0.2ex]	
$P_0$ & Initial Phage Dose & 100 & count  \\ [0.2ex] 
$A_0$ & Antibiotic Dose & 100 & count  \\ \hline
$H_{B}$ & Base Health** & $100$ & health units  \\
$R_{a}$ & Base Antibiotic Resistance (0 or 1) & 0 &   \\ 
$E_{B}$ & Base Energy** & $100$ & energy units  \\ [0.3ex]
$c_{met}$ & Base Metabolic Cost & 0.5 & energy units /tick  \\ [0.3ex]
 $r_{BF}$ & Base Adhesion Rate & 0.1 & count/interaction  \\ \hline
 $r_S$ & Nutrient Generation Rate & 300 & count/tick \\
  $t_A, \tau_A$ & Antibiotic Application Time / Dose Period & 60, 960 (8) & ticks  \\ 
$\tau_A$ & Antibiotic Dose Period & 960 (8) & ticks (hours) \\ 
 $\phi$ & Adsorption Rate & 0.005 & count/interaction  \\
 $r_{deg}$ & EPS Degradation Rate & 0.25 & count/interaction \\ 
 $N_{act}$ & Neutrophil Activation Threshold & 200 & count  \\ 
 $r_{upt}$ & Neutrophil Phagocytosis/Uptake Rate & 0.2 & count/interaction  \\          
\hline \hline
\end{tabular}\label{table2}
\end{center}
\end{table*}



\section{Results:} The objective of our simulation experiments is to better understand the effect of environmental factors such as temperature and pH on combined treatments of biofilm in both immunocompetent and immunocomproised patients. The NetLogo i\textit{n-silico} experiments have the potentials to simulate \textit{in-vivo} scenarios. The key variables in this experiment are the biofilm, antibiotics, phages and immunity which the key parameters are the temperature and the pH. These parameters are varied and we observe the optimal values that leads to better treatment of  the biofilm. We will first look at the baseline solutions for generational clusters and the effect of bacterial population density on the treatment dynamics.
Throughout this research endeavor, we employed the NetLogo programming environment as a powerful tool for simulating intricate biological processes. To establish a foundational dataset for our investigation, we chose to start with conducting a pathogens-only simulation. This approach allowed us to exclusively focus on the behavior of pathogens, enabling subsequent comparisons with other variables and experimental setups. A fundamental aspect of this study centers on comprehending the effects of diverse treatments in relation to the predefined generation clusters.

To visually depict the data, we designed the left-hand side display in Figure 1 to present generational growth curves. To facilitate line comparisons, generations were grouped into sets of five, each corresponding to a distinct color on the plot. The results presented are based on an average of 25 trials. Our analysis revealed a notably larger proportion of Generations 5 to 9 within the pathogen population. This observation signifies that substantial growth transpires during these generations. However, it is noteworthy that the deleterious impact of population density on the pathogen's division capability severely constrains the growth of subsequent generations.

\begin{figure}[!h]
\centering
\includegraphics[scale=.28]{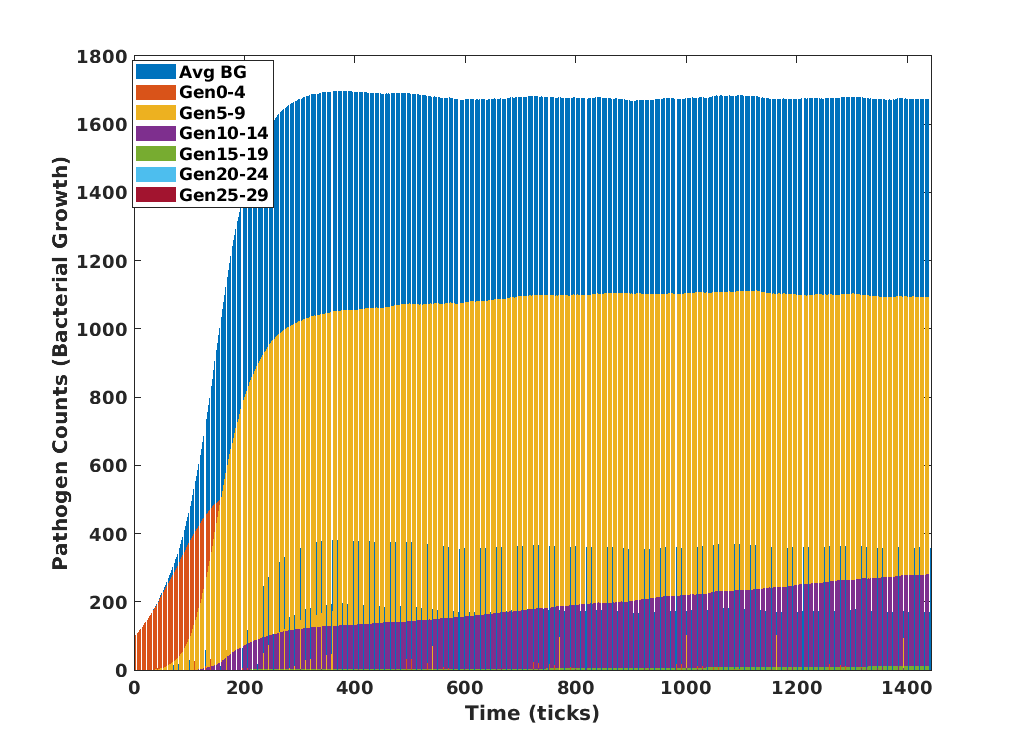} 
\includegraphics[scale=.33]{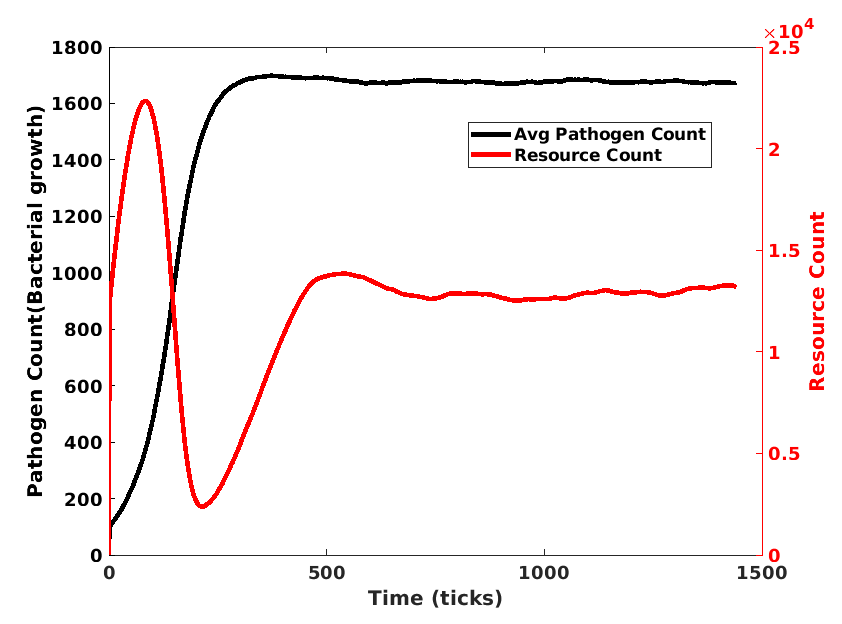}
 \caption{\textbf{Baseline Solution:: } From the left-hand figure, which represents the cluster of generations of pathogenic growth, is labeled by descending rainbow order to indicate the progression of clusters.  For example, the red color corresponds to Generations 0-4, orange is corresponds to Generations 5-9, and so forth. In the right-hand figure, the pathogen and resource count are graphed to provide a comparative look at their development throughout the simulation iterations}
    \label{fig1}
\end{figure}

On the right-hand side of Figure (\ref{fig1}), we present a comparative analysis of pathogen and resource counts, with the resource count representing the total number of nutrients distributed across all patches. The outcomes indicate that both populations reach a consistent level at approximately 400 ticks, equivalent to 200 minutes of simulated time. At this stage, spatial constraints and resource availability emerge as limiting factors for pathogen proliferation. A notable dip and subsequent rise in the resource count, observable between the one and two hundred tick marks, can be attributed to the stalling of pathogen growth due to spatial limitations. Subsequent mutations of the pathogens occur during this phase, as they tend to adapt and reduce their energy requirements for various physiological functions.

The comparative analysis of pathogen and resource counts shed light on the constraints imposed by spatial and resource limitations, offering valuable insights into pathogen adaptation in response to changing environmental conditions. These findings contribute to our understanding of the intricate dynamics governing pathogen populations, laying the groundwork for investigating the effects of various treatments on pathogen growth and behavior.



\subsection{Temperature and pH affects biofilm treatment \textit{in-vivo} \textit{in-silico} }



In this set of simulations results, we present a comprehensive investigation into the impact of temperature and pH levels on bacterial growth, based on ten simulation iterations. The y-axis of the heatmap represents various temperature values measured in Celsius, while the x-axis depicts pH values, collectively defining the environmental conditions under examination.

In the left-hand heatmap, each color on the legend corresponds to a specific numerical value representing the maximum population size attained at a given initial pH value and temperature combination. This representation effectively highlights the optimum conditions for bacterial growth within our simulation, enabling comparisons with other variations of these environmental factors.

On the right-hand plot, we present the average growth rate (measured in ticks) observed over ten simulation iterations. This growth rate denotes the time taken for the bacterial population to reach ninety-five percent of its maximum level in each iteration. We opted to distinguish and analyze these values to account for any abrupt changes in population development. Once the population reaches ninety-five percent of its maximum capacity, growth tends to stabilize, barring a few exceptional cases.

\begin{figure}[!h]
\centering
\includegraphics[scale=.31]{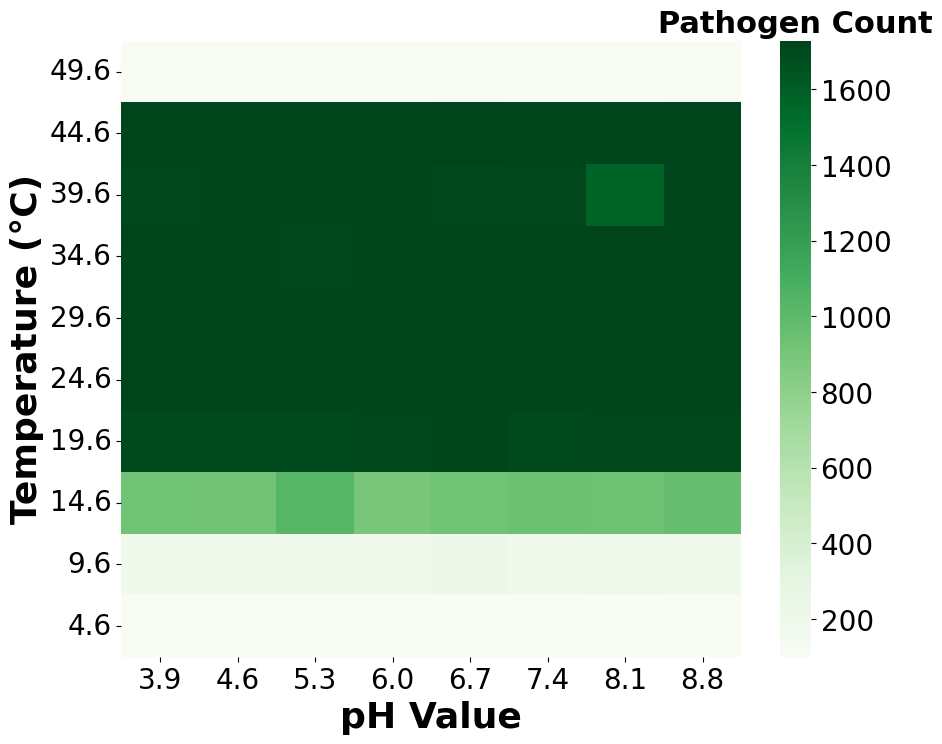} \ \ \
\includegraphics[scale=.31]{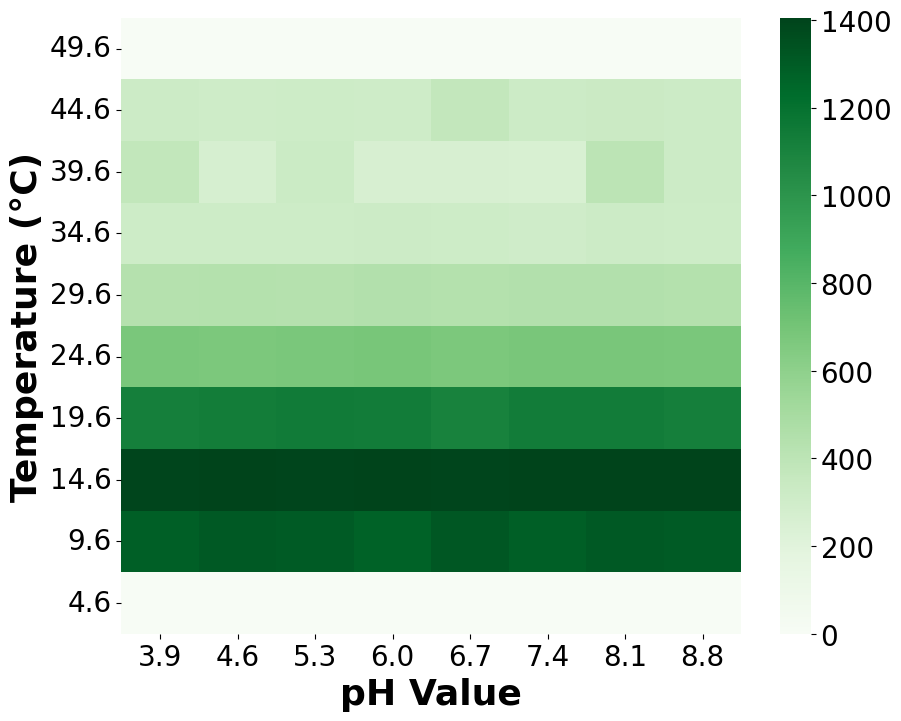}
 \caption{\textbf{Heatmaps for Temperature/pH - dependent Pathogen Count :} This figure shows the combined effect of temperature and pH value on the bacterial growth over $10$ simulations. The color bar from dark green to light green shows the bacterial count from high to low. The left plot is maximum population, while the right plot is for 95\% rise time}
    \label{fig2}
\end{figure}

From the heatmaps presented in Figure(\ref{fig2}), Each pair of heat maps represents the maximum pathogen population reached (left) and the average pathogen population (right) over the course of the simulation for varying initial temperature and pH values. The x-axis represents pH values, while the y-axis shows temperatures in degrees Celsius. Color corresponds to population value; a darker color indicates a larger value. Each coordinate on each heat map is the result of an average of 25 trials at those settings. For all trials, we assume free-floating bacterial pathogens, an immunocompetent setting, an application time of 60 ticks for both treatments, and a dose of 50 agents for both treatments. For treatments involving antibiotics, a second dose is administered 960 ticks after the initial dose. Overall, it shows that the effect of treatment can be seen even as low as 10 degree Celsius and 3.9 pH and up to 8.8 pH value. These heatmaps and growth rate plots provide valuable insights into the interplay between temperature, pH levels, and bacterial growth dynamics, thus contributing to a deeper understanding of the complex interactions governing microbial populations under varying environmental conditions.

\subsection{Single treatment depends on dosage rather than application time}
\begin{figure}[h!]
\centering
       \centering
\begin{tabular}{|l l | l l|}
       \hline
\multicolumn{2}{|c|}{Immunocompetent host} & \multicolumn{2}{|c|}{Immunocompromised host} \\
\hline
  \includegraphics[scale=.14]{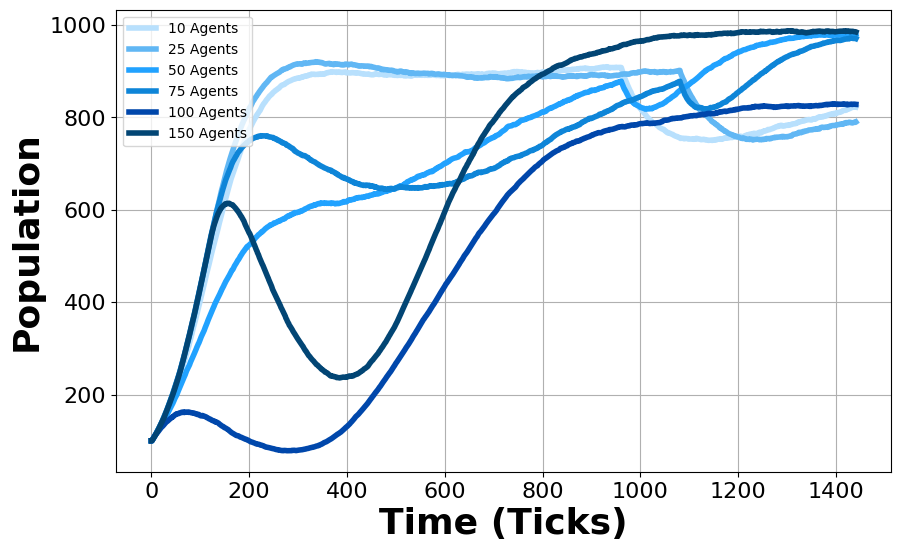}   & \includegraphics[scale=.14]{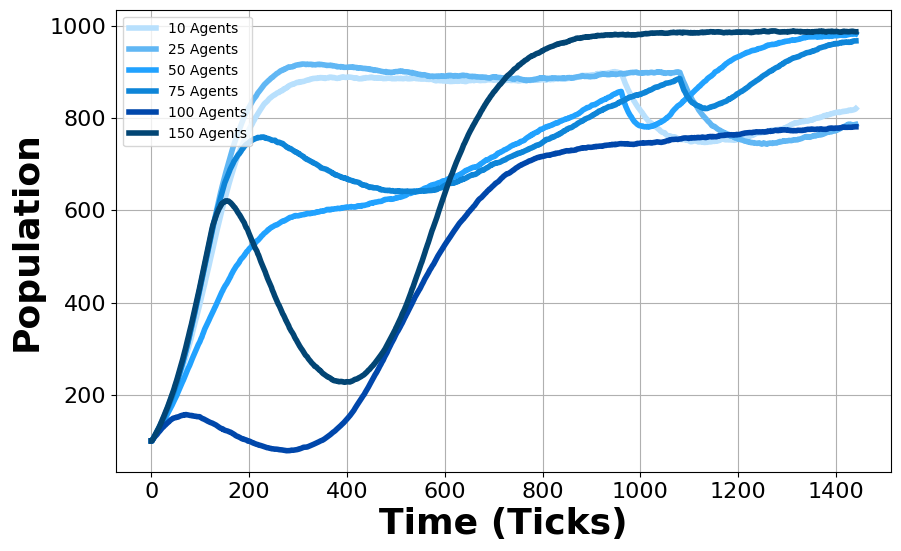} & \includegraphics[scale=.14]{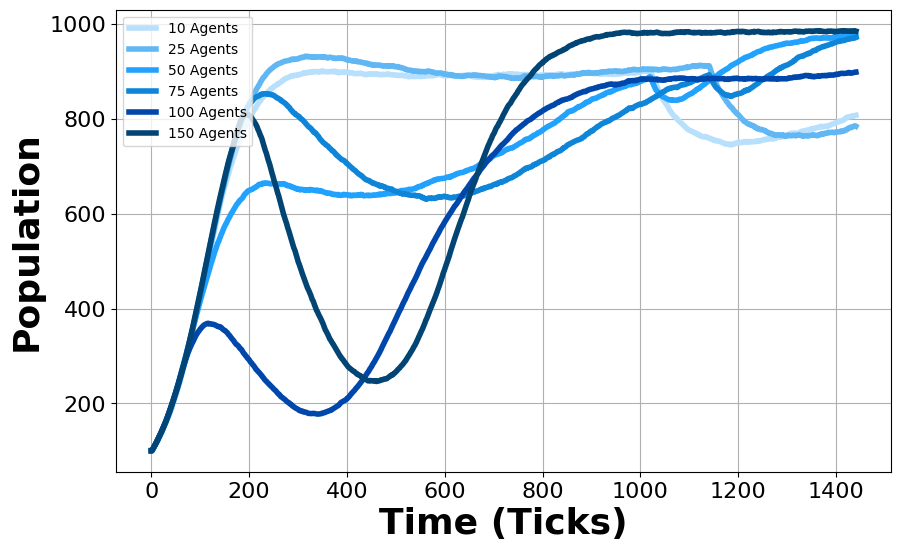}  & \includegraphics[scale=.14]{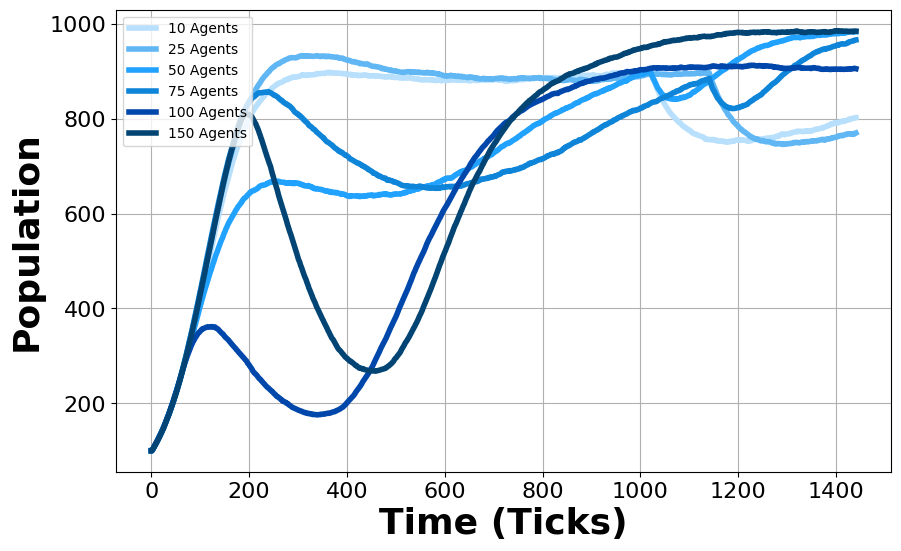}   \\
\includegraphics[scale=.14]{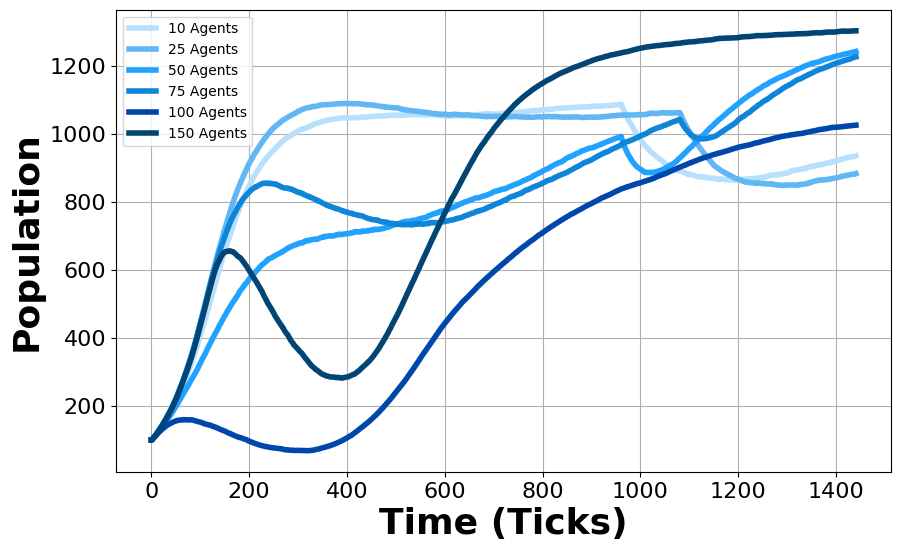}   & \includegraphics[scale=.14]{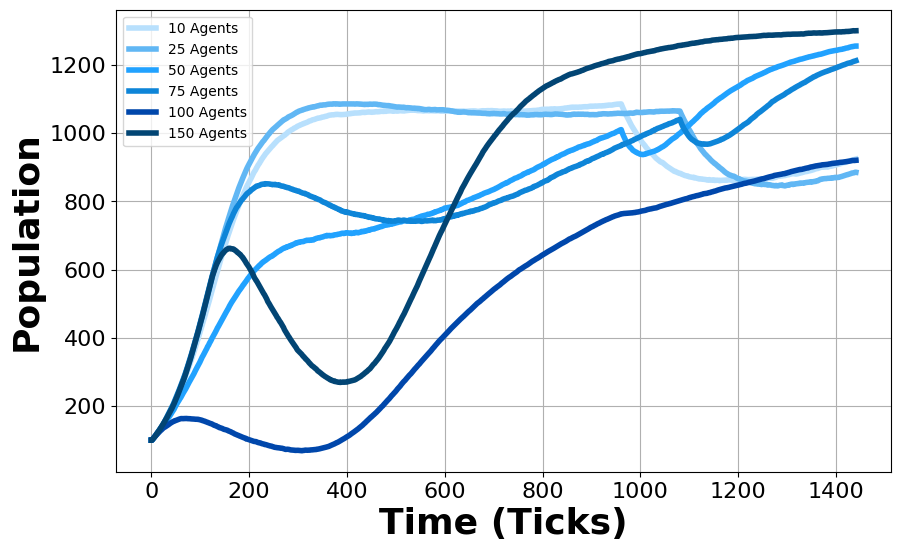} & \includegraphics[scale=.14]{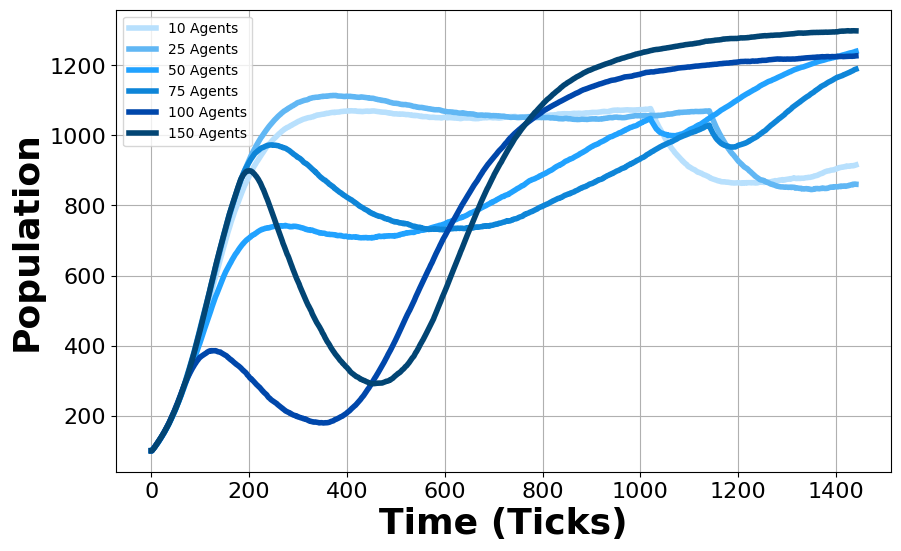}  & \includegraphics[scale=.14]{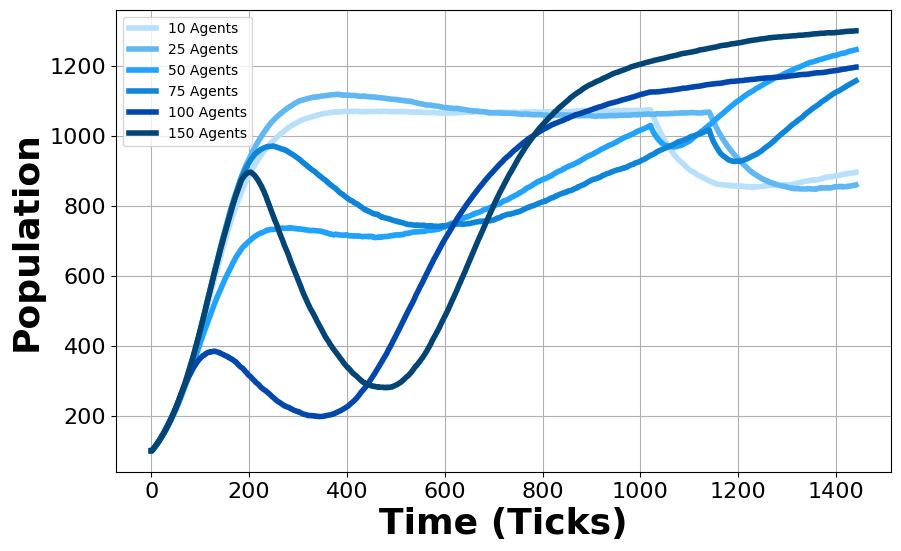}   \\ \hline
       \end{tabular} \caption{\textbf{Antibiotics-only simulations for immunocompetent and immunocompromised:} This set of simulations are at our four dose times being at 0, 60, 120, and 180 ticks. On the top left of both hosts, it features the initial dose application, followed by a second dose at 960 ticks.  On the top right of both hosts, it has an initial dose time of 60 ticks and a second dose at 1020 ticks.  The bottom left of both hosts features an initial dose time of 120 ticks and a second dose at 1080 ticks.  Lastly, the bottom right of both hosts has an initial dose time of 180 ticks and a second dose time of 1140 ticks.  Each of the individual lines on the four plots of both hosts represents the number of antibiotic agents introduced. The lighter the color, the fewer agents, in the order of 10, 25, 50, 75, 100, and 150.}\label{fig:antibiotics}
 \end{figure}

We observed that the application dose of antibiotics or phages has more effect on the treatment of bacterial biofilm than the application time, irrespective of the immunity status of the host. So, in the first case, we consider the treatment of bacterial pathogens using only antibiotics. We explore both the immunocompetent and immunocompromised settings, which include and exclude innate immune cells (neutrophils) respectively. We assume the pathogens to be free-floating. We recorded the population over time for a total of 1440 ticks (12 hours) for varying combinations of antibiotic application times and doses and for each of the immunocompetent and immunocompromised settings, displayed in Figure (\ref{fig:antibiotics}). A series of 25 trials was conducted at each of 4 different application times ranging from 0 to 180 ticks (1.5 hours) and 6 different doses ranging from 10 to 150 agents.



\begin{figure}[h!]
\centering
       \centering
\begin{tabular}{|l l | l l|}
       \hline
\multicolumn{2}{|c|}{Immunocompetent host} & \multicolumn{2}{|c|}{Immunocompromised host} \\
\hline
  \includegraphics[scale=.14]{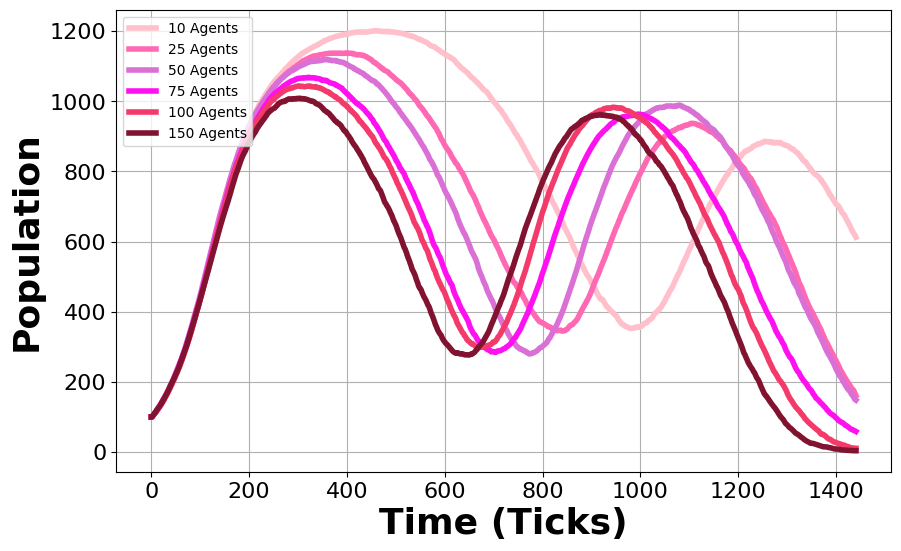}   & \includegraphics[scale=.14]{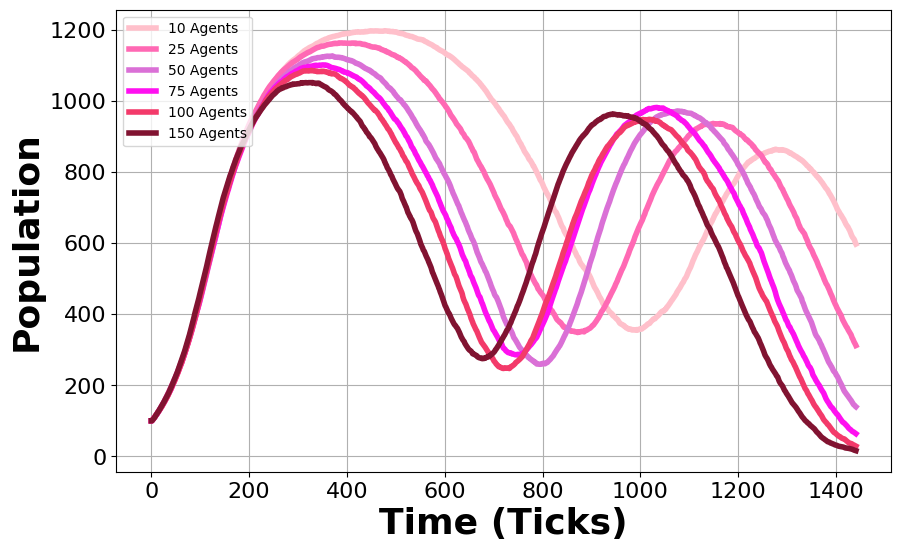} & \includegraphics[scale=.14]{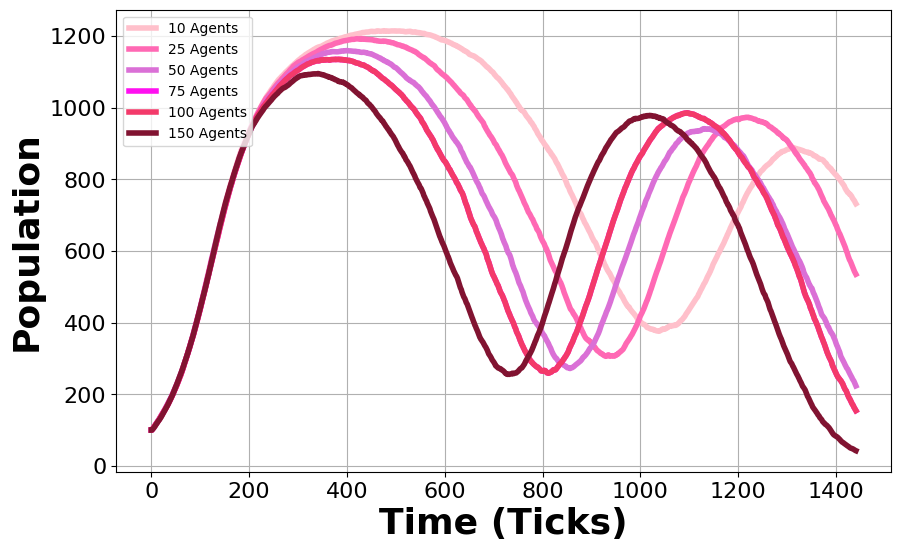}  & \includegraphics[scale=.14]{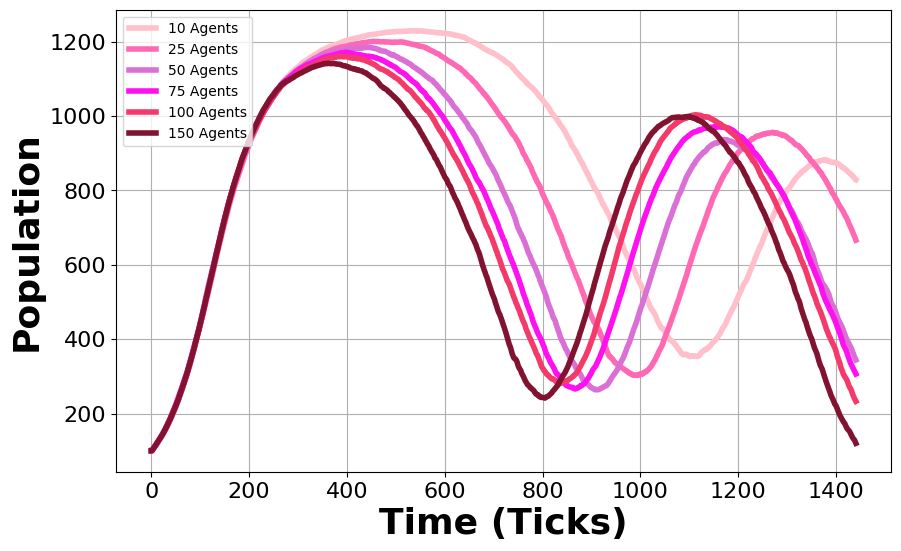}   \\
\includegraphics[scale=.14]{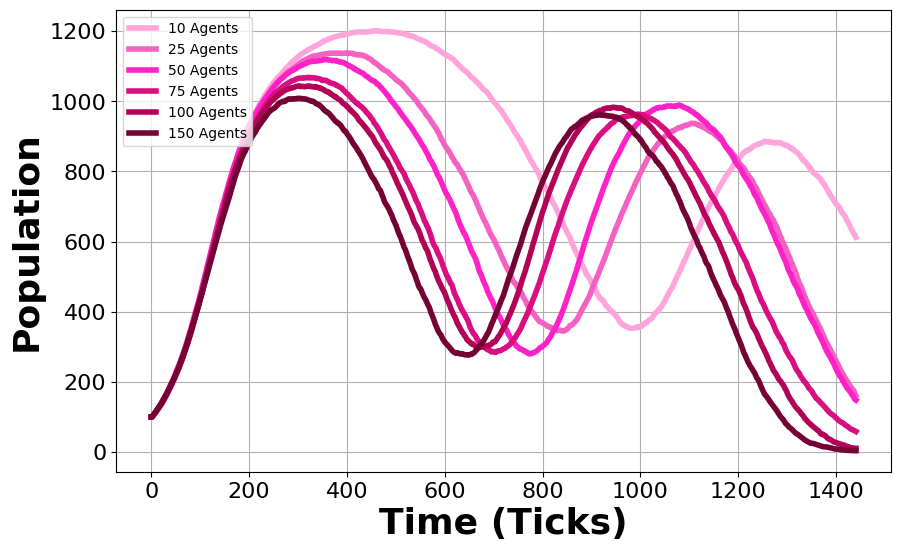}   & \includegraphics[scale=.14]{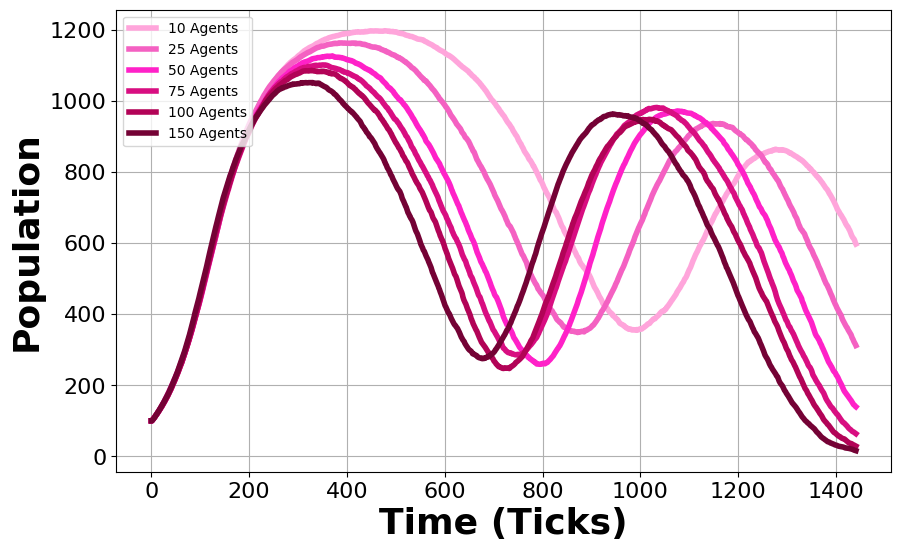} & \includegraphics[scale=.14]{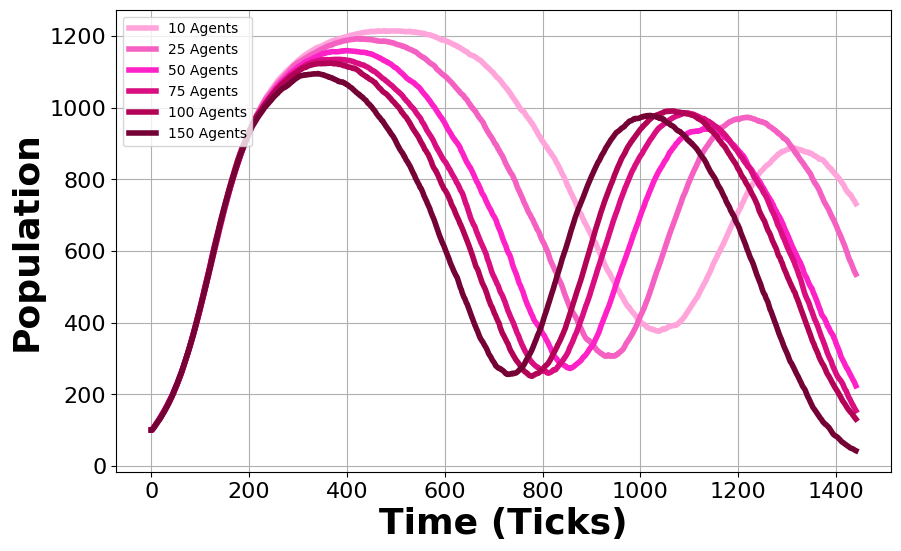}  & \includegraphics[scale=.14]{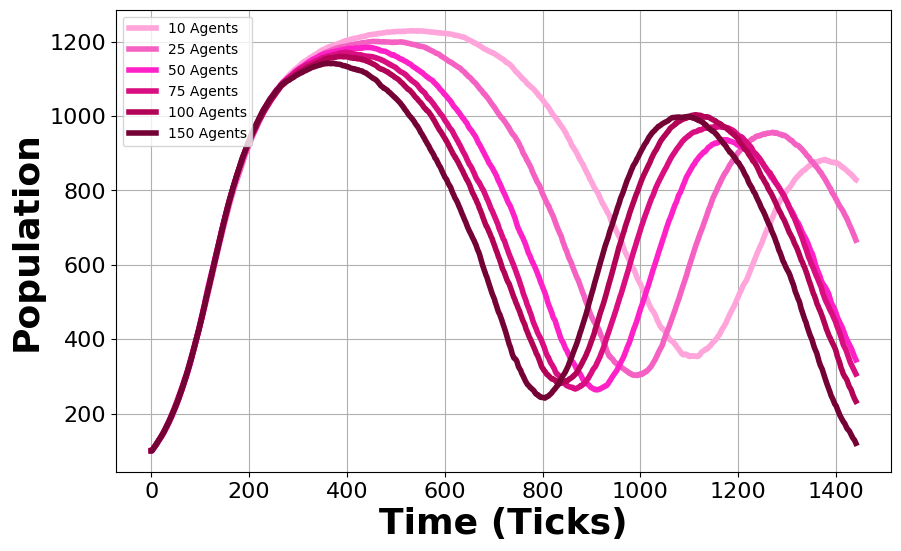}   \\ \hline
       \end{tabular} \caption{\textbf{Phages-only simulations, immunocompetent and immunocompromised:} 
       Presented in this figure are the results for the treatment of free-floating pathogens with only phages in both immunocompetent and immunocompromised settings with varying application times and doses. On the left box, the four plots represent the results from the mmunocompetent simulations, while the right box represents the immunocompromised results. Going from the left-hand side to the right-hand side of each box (host), the columns are in order from a 0 tick dose time, 60 tick dose time, 120 tick dose time, and 180 tick dose time. On the x-axis, the time indicates how far the simulation has progressed.} \label{fig:phage}
 \end{figure}

In the case of a single treatment with antibiotics, we observe that for small doses (10 and 25), pathogens are not forced to evolve resistance and thus remain susceptible to antibiotics for the second dose, while for larger doses (100 and 150), the population dips drastically and the activity of the antibiotics forces pathogens to evolve resistance. These resistant pathogens can rebound to the original full population and can even ultimately succeed the smaller doses in growth. Interestingly, the medium doses (50 and 75) are too small to eliminate all susceptible pathogens but still drive some evolution of resistance. In addition, this selective pressure can drive other mutations which benefit long-term growth, which can be seen in the ultimate return to a high population level after the second dose. Application time does not appear to greatly affect the response. The lack of an immune response does not greatly affect the shapes of the curves, but it does allow the pathogens to reach roughly 20\% more total growth across the duration of the simulation

In the case of a single treatment with phages, we consider the treatment of bacterial pathogens using only phages. We explore both the immunocompetent and immunocompromised settings, which include and exclude innate immune cells (neutrophils) respectively. We assume the pathogens to be free-floating. We recorded the population over time for a total of 1440 ticks (12 hours) for varying combinations of phage application times and doses and for each of the immunocompetent and immunocompromised settings, displayed in Figure(\ref{fig:phage}). A series of 25 trials was conducted at each of 4 different application times ranging from 0 to 180 ticks (1.5 hours) and 6 different doses ranging from 10 to 150 agents. The oscillations are the result of cyclic dynamic interactions with the predatory phage population. So, as the phages begin to grow and prey on the pathogens, the bacterial pathogen population dips while the phages exhibit exponential growth, this stalls as the pathogen population reaches a low value, between 600 and 800 ticks (depending on application time), allowing the pathogens to rebound. This rebounding allows the phages to begin growing exponentially once again, decimating the pathogen population towards the end of the simulation. We observe that a larger initial phage dose speeds up and heightens the oscillations, but it does not significantly affect the maximum pathogen population reached; resistance to phages tend to evolve in a minority of the pathogens, but only to the original phage genotype and not to any of the phages which mutated. This does not appear to be a major factor in the treatment dynamics, as the pathogens clearly remain susceptible to pathogen for the second period of population decline around 1200 ticks. Overall, we observe similar dynamics for both hosts, so the presence of immunity does not greatly affect the curve shapes or maximum pathogen growth. We will like to remark that some of the trials with larger doses of antibiotics and phages does significantly reduced the pathogen count in the system to almost zero, but this is not nicely reflected in Figure(\ref{fig:antibiotics}) and Figure(\ref{fig:phage}) respectively because we plotted the average of the 25 trials.


\subsection{Combined Treatments is induced by phage and antibiotic threshold}

\begin{figure}[!t]
\centering
\includegraphics[scale=.24]{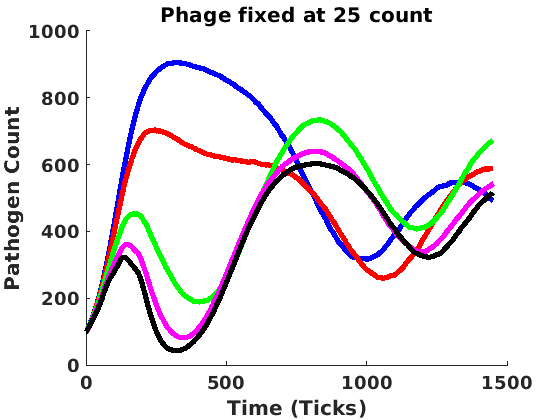}    
\includegraphics[scale=.24]{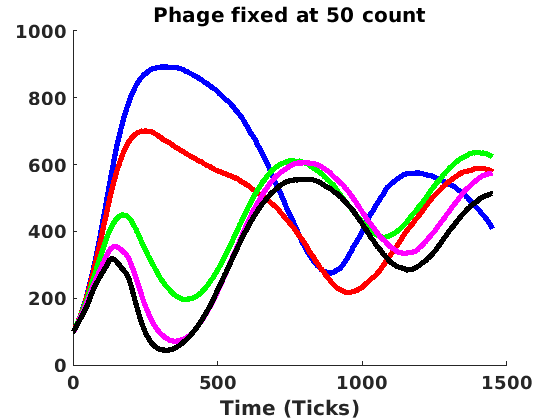} 
\includegraphics[scale=.24]{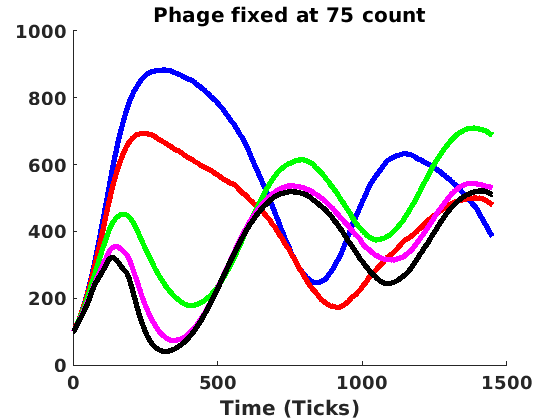} 
\includegraphics[scale=.24]{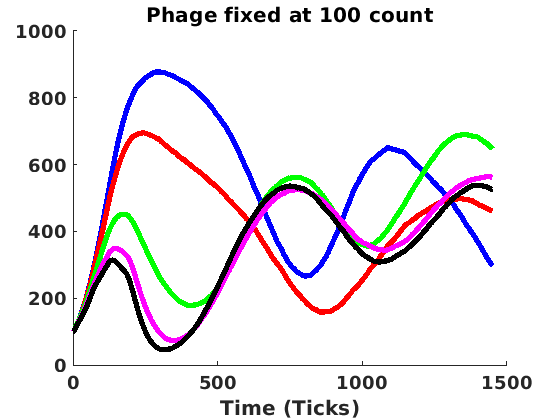} \\
\includegraphics[scale=.24]{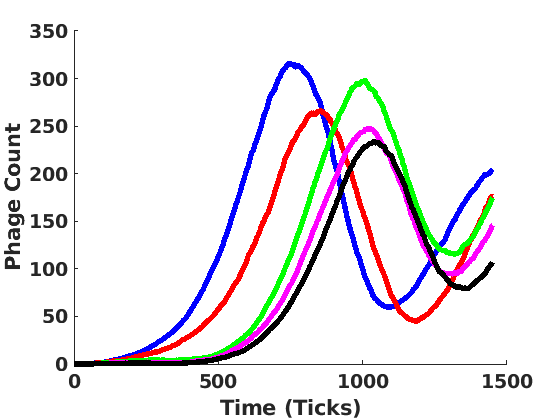}    
\includegraphics[scale=.24]{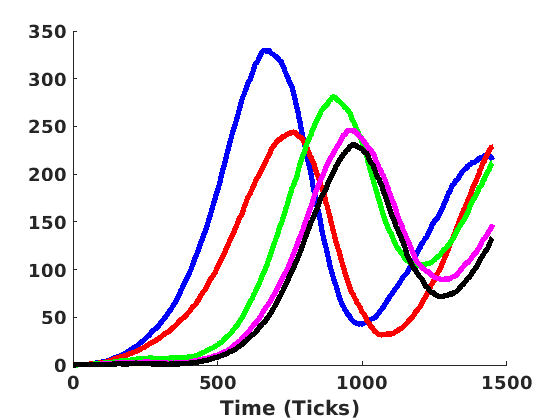} 
\includegraphics[scale=.24]{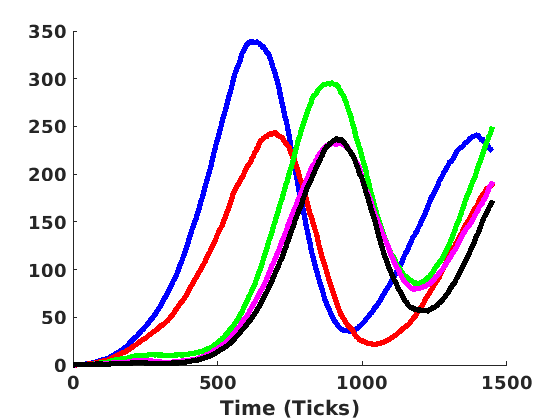} 
\includegraphics[scale=.24]{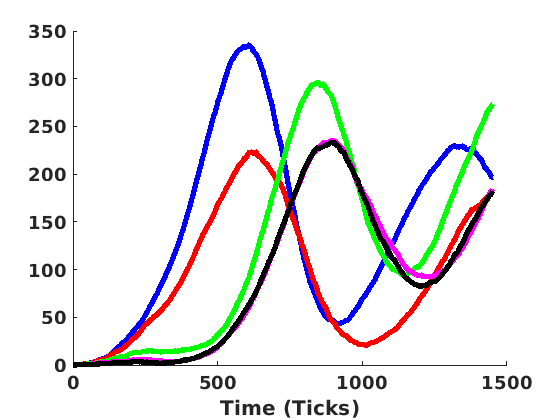} \\
\includegraphics[scale=.24]{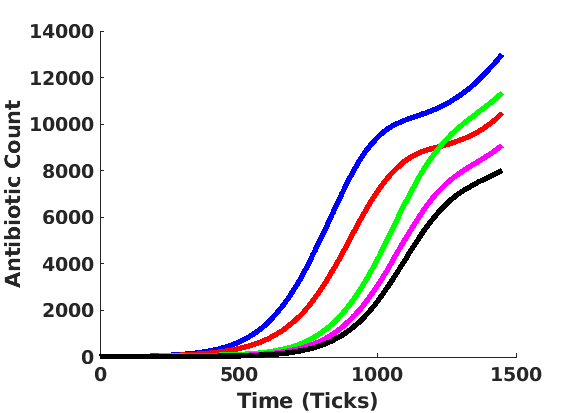}    
\includegraphics[scale=.24]{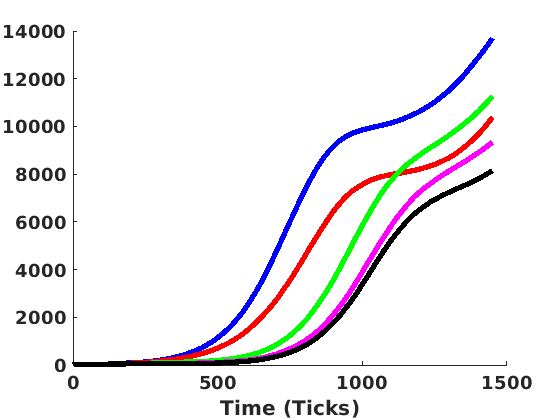} 
\includegraphics[scale=.24]{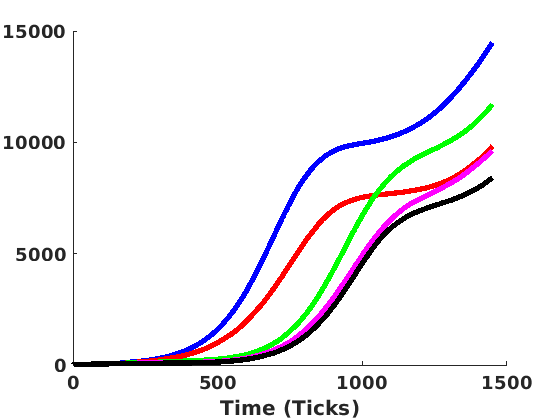} 
\includegraphics[scale=.24]{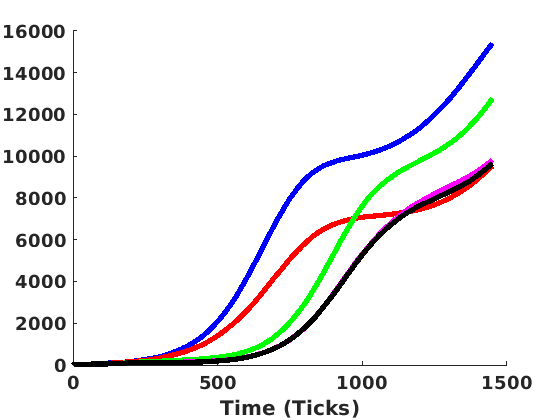} \\
 \caption{\textbf{Temporal plots of immunocompetent case with varying Antibiotic doses at fixed phage counts: } Here will varied the different doses of Antibiotics: $10, 25, 50, 75$ and $100$, applied at four specific times: $0, 60, 120$ and $180$ while fixing the initial phage count at $25, 50, 75$ and $100$ respectively for each of the simulations. The temporal plots are presented along the rows for the Pathogen Count, Phage Count and Antibiotic Count respectively}
    \label{fig:combined-immunocompetent}
\end{figure}
\begin{figure}[!t]
\centering
\includegraphics[scale=.24]{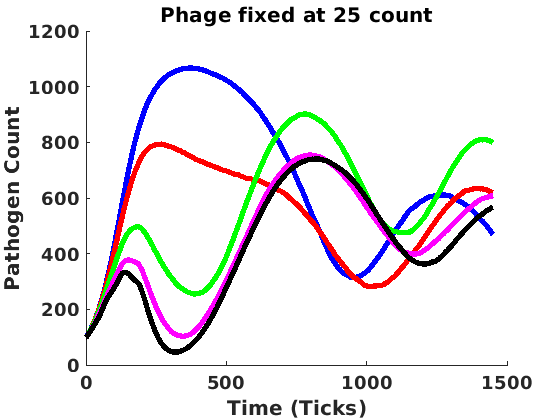} 
\includegraphics[scale=.24]{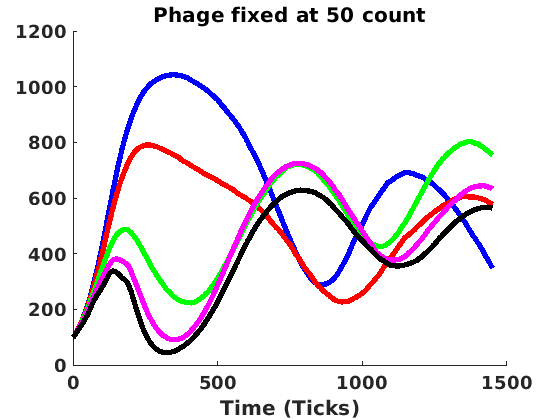} 
\includegraphics[scale=.24]{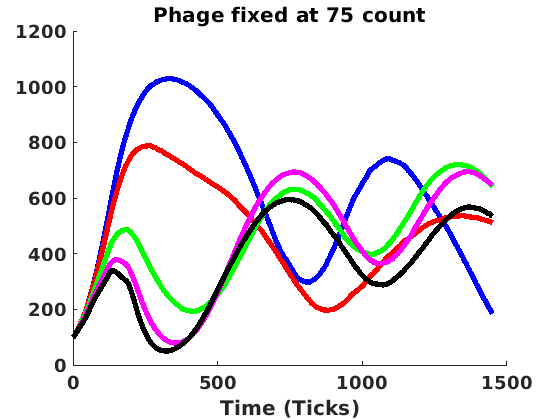} 
\includegraphics[scale=.24]{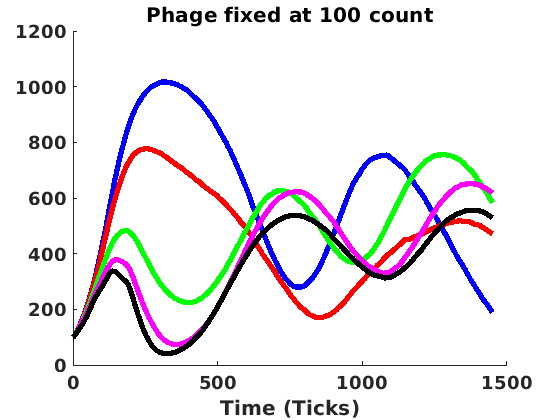} \\
\includegraphics[scale=.24]{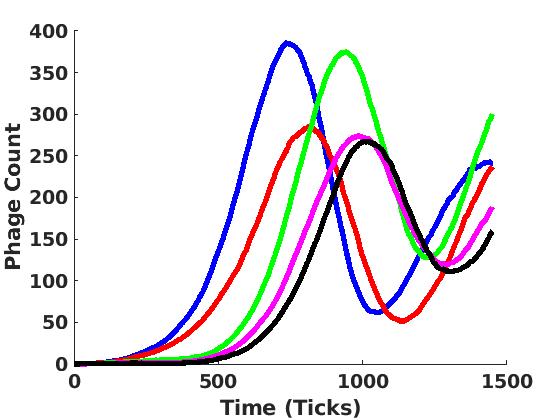}   
\includegraphics[scale=.24]{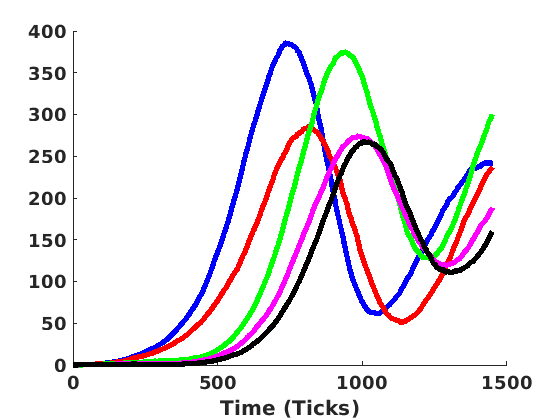} 
\includegraphics[scale=.24]{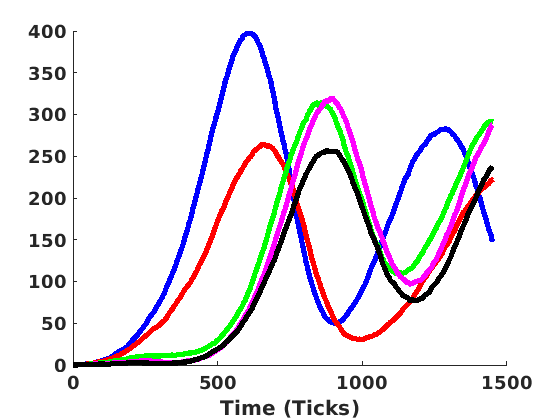} 
\includegraphics[scale=.24]{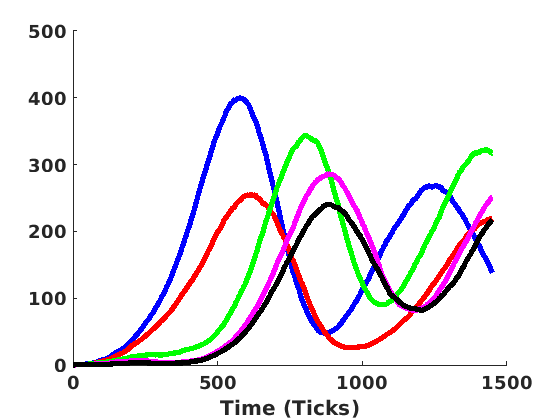} \\
\includegraphics[scale=.24]{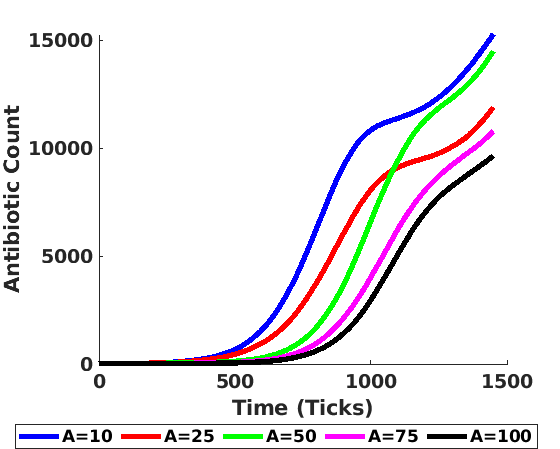} 
\includegraphics[scale=.24]{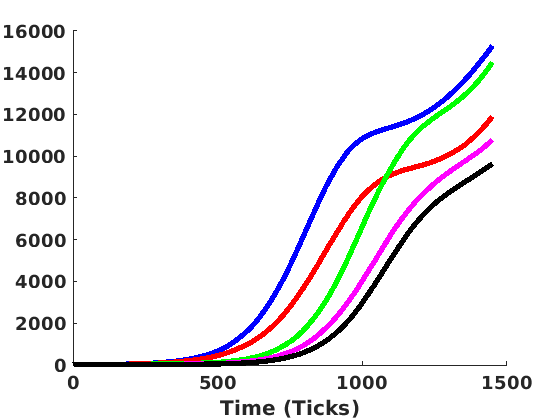} 
\includegraphics[scale=.24]{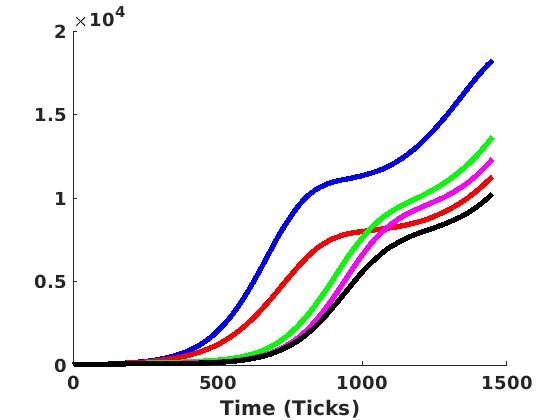} 
\includegraphics[scale=.24]{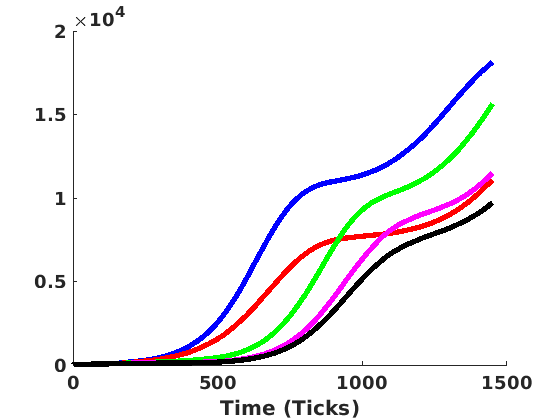} \\
 \caption{\textbf{Temporal plots of immunocompromised case with varying Antibiotic doses at fixed phage counts: } Here will varied the different doses of Antibiotics: $10, 25, 50, 75$ and $100$, applied at four specific times: $0, 60, 120$ and $180$ while fixing the initial phage count at $25, 50, 75$ and $100$ respectively for each of the simulations. The temporal plots are presented along the rows for the Pathogen Count, Phage Count and Antibiotic Count respectively} 
    \label{fig:combined-immunocompromised}
\end{figure}

The efficacy of combined antibiotic and phage therapy for treating bacterial pathogens has been explored under varying conditions of host immunity. This study investigates the effects of such treatments in both immunocompetent and immunocompromised settings, as depicted in Figure(\ref{fig:combined-immunocompetent}) and Figure(\ref{fig:combined-immunocompromised}) respectively. The pathogens in these experiments are assumed to be free-floating, and their population dynamics are recorded over a period of 12 hours, or 1440 ticks, under different conditions of application timing, antibiotic doses, and phage doses. The study involved 25 trials for each combination of parameters, with application times ranging from 0 to 180 ticks (1.5 hours) and doses varying from 10 to 150 agents. The focus was on understanding the interaction between antibiotics and phages in controlling pathogen populations, particularly under different immune statuses of the host. In both immunocompetent and immunocompromised settings, the combined treatment of antibiotics and phages resulted in significantly less overall pathogen growth compared to single treatments. The data suggest that a high initial dose of treatment agents and a quick application time are crucial for sustaining low pathogen growth rates. This effect is particularly noticeable with a threshold phage count of 50 or more and an antibiotic dose of 100 or more, which consistently led to a reduced pathogen count and infection level. A notable observation was the sudden decline in pathogen count at the second application of antibiotics, especially when initial antibiotic doses were low and phage doses were high. This suggests that the timing and dosage of treatment play critical roles in the success of the combined therapy. Specifically, high antibiotic doses $(A > 50)$ in conjunction with phage counts exceeding 50 consistently reduced infection levels, highlighting the importance of adequate dosing. Interestingly, the decrease in pathogen count was not always mirrored by an increase in phage growth, particularly at initial application times. This indicates that the treatment's efficacy is not solely dependent on phage proliferation but rather on the synergistic action of both antibiotics and phages. The phages' role seems to be more about maintaining low pathogen levels rather than solely relying on their replication to achieve therapeutic outcomes. At lower antibiotic doses, some level of pathogen resistance was observed, as indicated by higher pathogen counts. However, this scenario also provided an opportunity for phages to target and reduce the pathogen population, as evidenced by a subsequent increase in phage counts. This interaction eventually led to a degradation of the pathogens and a reduction in infection levels, especially in immunocompromised settings, as shown in Figure(\ref{fig:combined-immunocompromised}). The results of this study demonstrate that the combined use of antibiotics and phages can effectively control bacterial pathogens, with the treatment's success heavily influenced by the timing and dosage of the agents used. The findings underscore the importance of an integrated approach, leveraging both antibiotics and phages to achieve optimal therapeutic outcomes. While the decrease in pathogen count was not always accompanied by increased phage growth, the synergy between the two treatments proved crucial in managing infections, particularly in cases where host immunity is compromised. This research contributes valuable insights into the design of effective treatment strategies against bacterial infections, especially in the context of rising antibiotic resistance. 

\begin{figure}[h!]
\centering
       \centering
\begin{tabular}{|l| l|}
       \hline
Low Antibiotic $A=10$ & High Antibiotic $A=150$ \\
\hline
  \includegraphics[scale=.45]{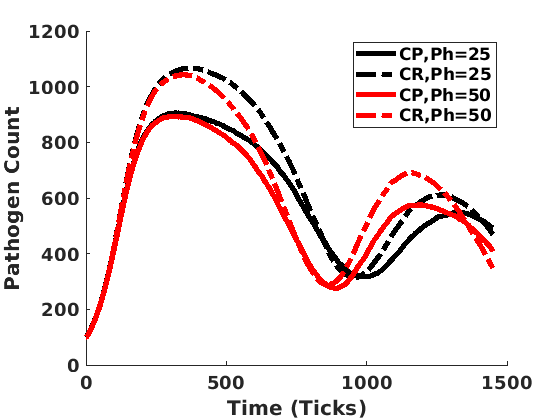}  &
\includegraphics[scale=.45]{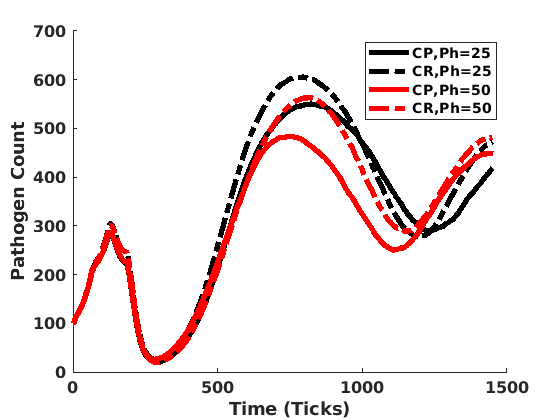} \\
\includegraphics[scale=.45]{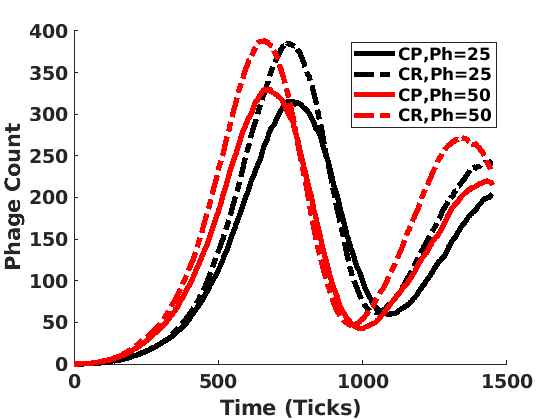} &
\includegraphics[scale=.45]{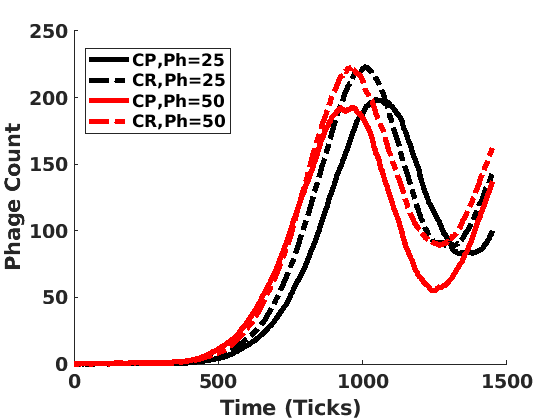}   \\ \hline
       \end{tabular}  \caption{\textbf{Temporal plots comparing pathogens and the phages for immunocompetent and immunocompromised cases with low and high antibiotics doses: } Here we varied Antibiotics between low and high dosage, $10$ and $150$ respectively, while fixing the initial phage count at $25$ and $50$ respectively for each of the simulations for imuunocompetent and immunocompromised settings respectively. From the legend, Ph represents fixed Phage count, CP represents immunocompetent, while CR represents immunocompromised. The first column is for low antibiotics and the second column is for high antibiotics. The top row is for pathogen count while the bottom row is for phage count.}     \label{fig:comparing-agentshighlow}
 \end{figure}

Our findings is further confirmed by the results presented in Figure(\ref{fig:comparing-agentshighlow}), whereby we compared fixed phage counts with low and high antibiotics. Overall, a combination of fixed phage count with high antibiotics has more effect on the immunocompetent case than the immunocompromised case. 


\subsection{Biofilm treatment controls biofilm development and structure}

Notice that so far we have been looking at the effect for free-floating bacteria. In this subsection, we will take the biofilm structure and properties into account. Starting treatment at the begining of biofilm formation does not provide any interesting information, since the expected outcome can be easily predicted. We would rather prefer to treat a growing biofilm instead of waiting till it is matured, hence we will run a simulation of both single and combined treatments on a growing (medium stage) biofilm. For each of the treatment strategies, we will monitor the treatment effects at the onset of treatments and three other time intervals as show in Figure (\ref{fig:biofilmtreat}). 

\begin{figure}[!h]
\centering
\includegraphics[scale=.2]{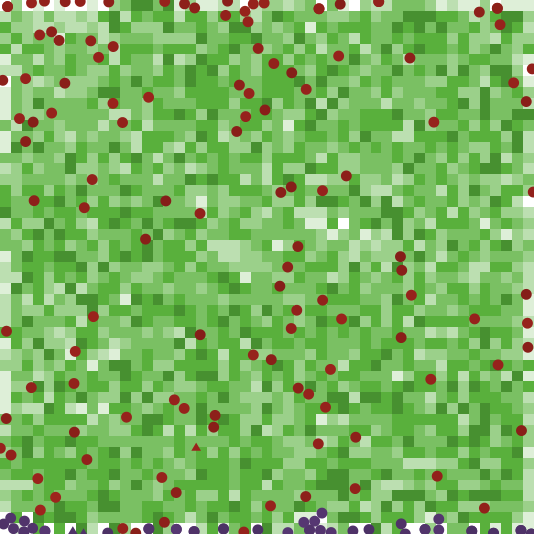} \ \ \
\includegraphics[scale=.2]{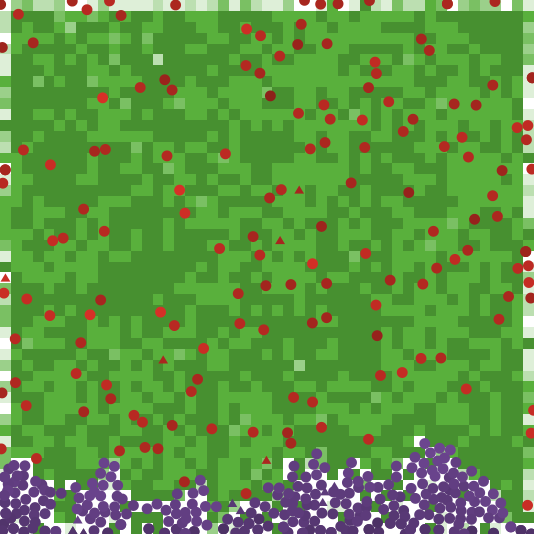} \ \ \includegraphics[scale=.2]{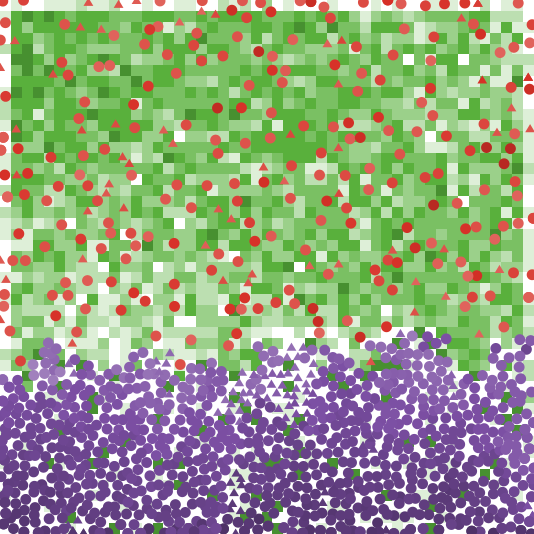}
 \caption{\textbf{The three stages of biofilm development  with Pathogens only: } The simulation from left to right represents early, growing, and mature biofilm. These snapshots were taken from a simulation with only pathogens starting with pathogen of 45 counts, growing biofilm is at 300 counts while the matured biofilm is at 1000 pathogen counts. biofilm cells respectively. The red icons represent planktonic (free-floating) pathogens, purple icons represent sessile (biofilm) pathogens, blue crosses represent antibiotics, and black dots represent phages. The antibiotics and phages are not necessarily attacking the pathogents. Circles represent pathogens who have not undergone a mutation, while triangles represent pathogens who have. Lighter pathogen colors represent younger generations. Green squares represent patches containing nutrients, with darker greens representing greater concentrations of nutrients.}
    \label{fig:biofilm-development}
\end{figure}

\begin{figure}[h!]
\centering
       \centering
\begin{tabular}{|c |l l  l l|}
       \hline
& Initial & After 4hrs & After 8hrs & After 12hrs \\\hline

 \rotatebox[origin=l]{90}{\textbf{Phage Treatment}} & \includegraphics[scale=.2]{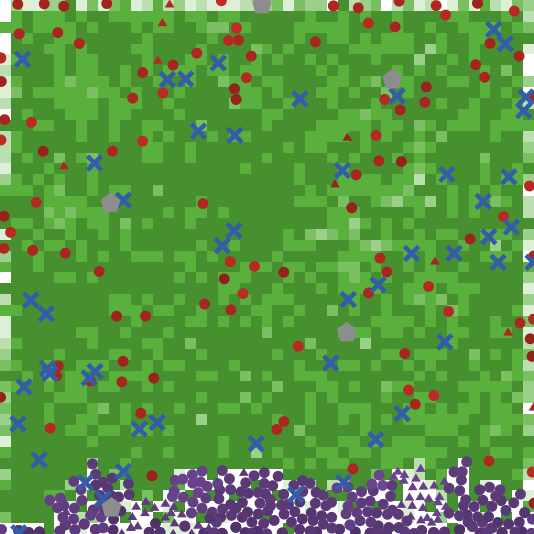} &
\includegraphics[scale=.2]{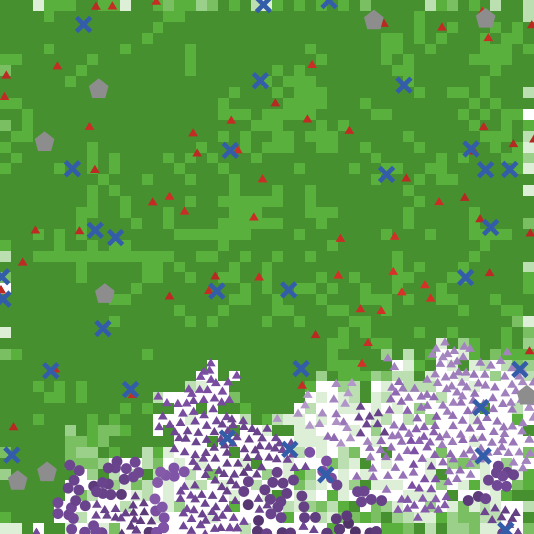} &
\includegraphics[scale=.2]{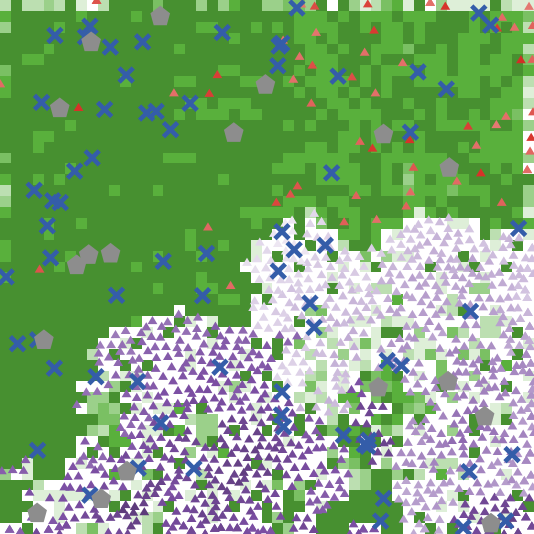} & \includegraphics[scale=.2]{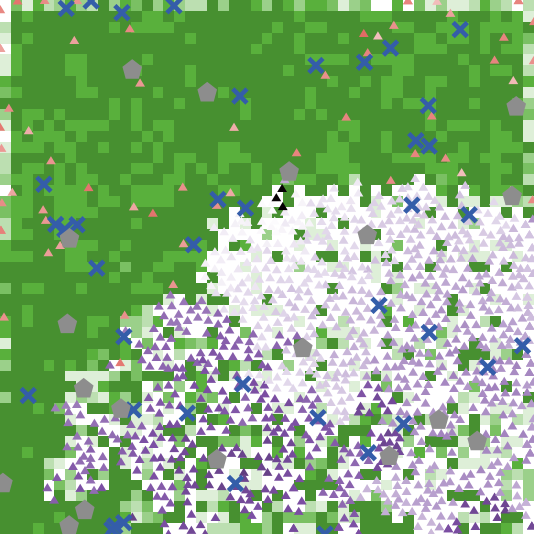}  \\\hline
 \rotatebox[origin=l]{90}{\textbf{Antibiotic Treatment}} &  \includegraphics[scale=.2]{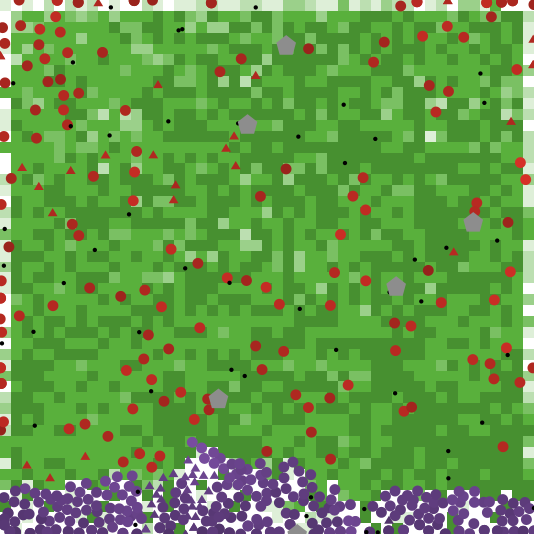} &
\includegraphics[scale=.2]{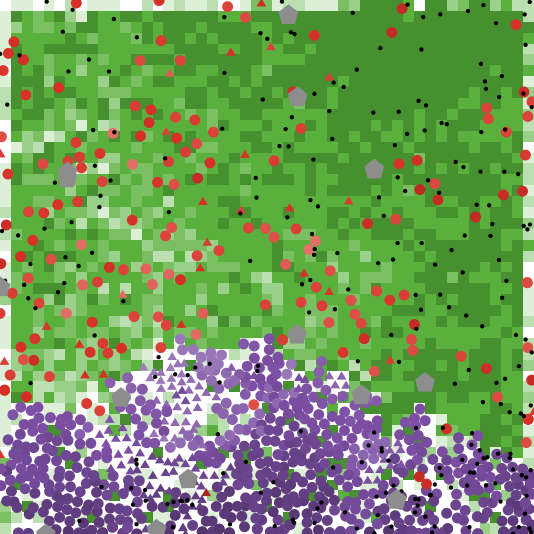} &
\includegraphics[scale=.2]{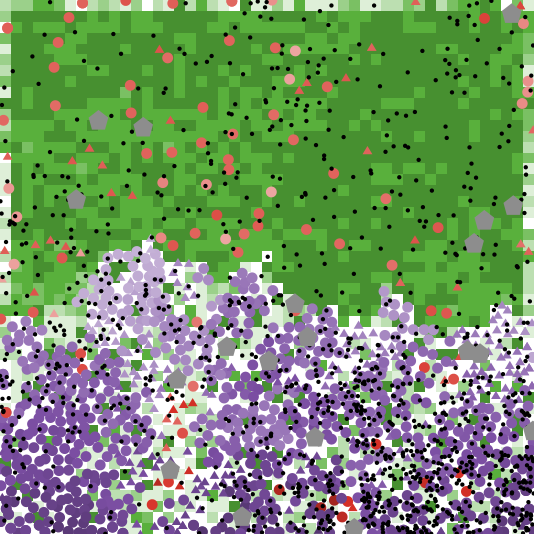} & \includegraphics[scale=.2]{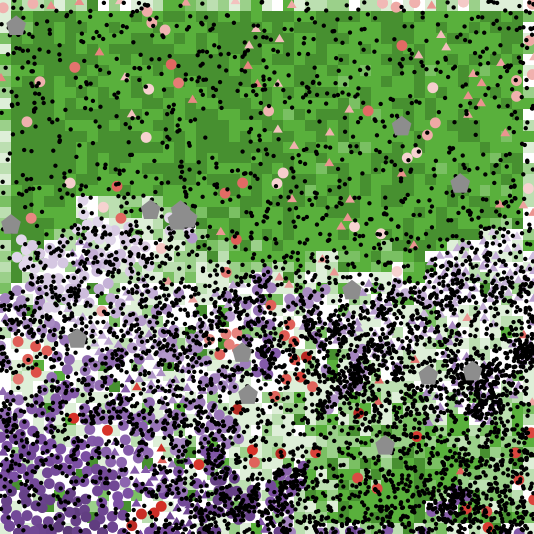}  \\\hline
 \rotatebox[origin=l]{90}{\textbf{Combined Treatment}} &  \includegraphics[scale=.2]{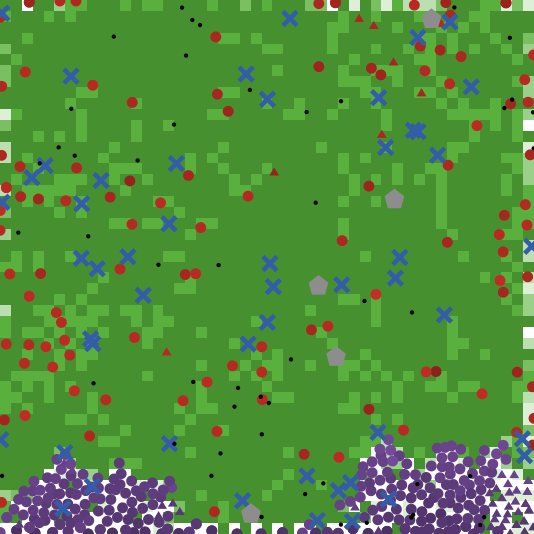} &
\includegraphics[scale=.2]{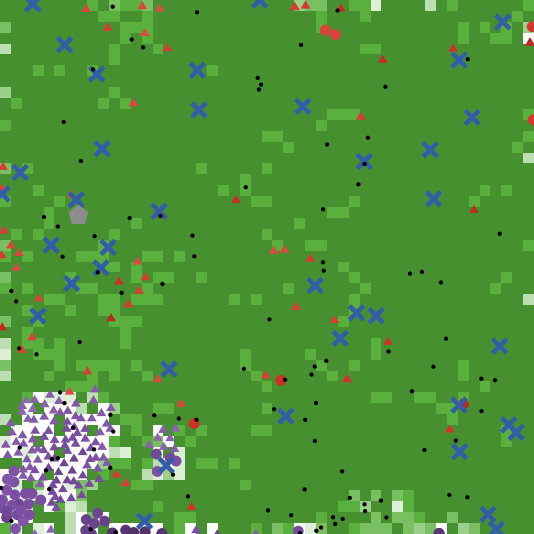} &
\includegraphics[scale=.2]{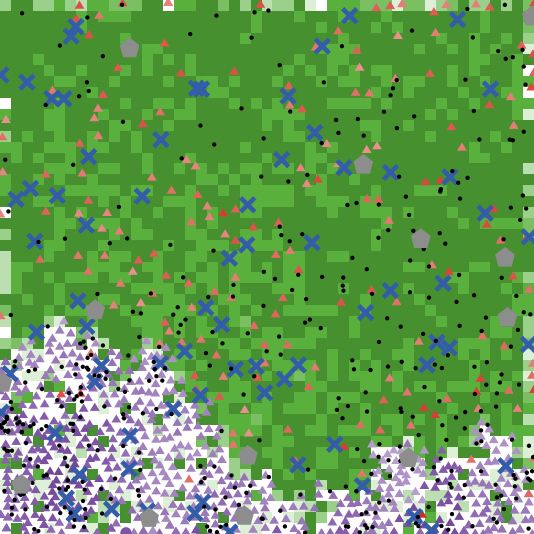} & \includegraphics[scale=.2]{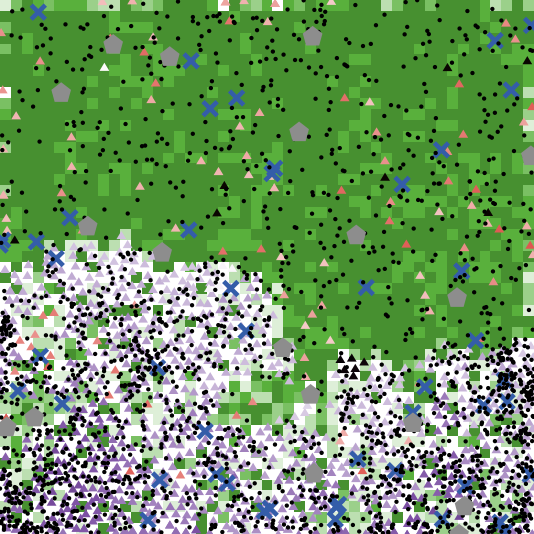}  \\ \hline
       \end{tabular} \caption{\textbf{Single and combined treatment in a growing biofilm:} 
       This figure represents biofilm degradation by single treatment (phage treatment on top row, antibiobiotic treatment on middle row) and degradation by a combination of antibiotics and phages on the bottom row. Snapshots are taken when treatment is applied (0 hrs), and after 4 hrs, 8 hrs, and 12 hrs respectively. These snapshots were taken from a simulation with antibiotic and phage doses of 50 counts administered at the growing stage of the biofilm.} \label{fig:biofilmtreat}
 \end{figure}

The formation of a biofilm through the three stages of development (early, growing and matured) is shown in Figure(\ref{fig:biofilm-development}). The development of biofilms is mostly not uniform across their surface. Various sections of the biofilm can grow at different rates, leading to an irregular, jagged appearance. Some regions may extend outward more significantly than others, resulting in a non-uniform biofilm border. This uneven growth can be attributed to several factors, including variations in nutrient availability, oxygen levels, and the local environment's physical and chemical conditions. Areas of the biofilm with better access to nutrients and oxygen tend to grow more rapidly, causing portions of the biofilm to jut out. Conversely, regions with limited access to these resources may exhibit slower growth or even stunted development. This differential growth can also be influenced by the nature of the surface to which the biofilm adheres, the hydrodynamics of the surrounding fluid, and interactions with other microorganisms. These factors collectively contribute to the heterogeneous architecture of biofilms, which can impact their overall function and resilience. Within biofilms, the distribution of cells is not homogenous. Newer cells are generally found towards the outer edge, while older cells are located deeper within the biofilm matrix. This stratification can be visually represented, with lighter shades indicating newer cells and darker shades denoting older ones. The high density of immobile pathogens in the deeper layers of the biofilm plays a crucial role in this spatial arrangement.

In the inner regions of the biofilm, cells often face nutrient and oxygen limitations due to their restricted access. These conditions hinder the ability of older cells to divide, resulting in a lower growth rate. As a consequence, the newer, more actively dividing cells are pushed toward the periphery, where conditions are more favorable. This phenomenon can create a dynamic gradient of cellular activity, with more metabolically active and proliferative cells located on the outer edges of the biofilm.

Within biofilms, the distribution of cells is not homogenous. Newer cells are generally found towards the outer edge, while older cells are located deeper within the biofilm matrix. This stratification can be visually represented, with lighter shades indicating newer cells and darker shades denoting older ones. The high density of immobile pathogens in the deeper layers of the biofilm plays a crucial role in this spatial arrangement.

The spatial structure of biofilms also contributes to the formation of clusters of mutant pathogens. As biofilm-associated cells divide, mutations can arise and be propagated within the population. Given the immobile nature of cells within the biofilm matrix, these mutated pathogens are often confined to a specific location. As they continue to divide and pass on their mutated properties, clusters of mutants form, leading to localized regions with distinct genetic and phenotypic characteristics as presented in Figure(\ref{fig:biofilm-development}).

This clustering of mutants is significant for several reasons. First, it can lead to the emergence of specialized subpopulations within the biofilm, potentially possessing enhanced resistance to antimicrobial agents or other stressors. Second, these mutant clusters can contribute to the overall genetic diversity of the biofilm community, increasing its adaptability and resilience. Finally, the presence of such clusters may pose challenges for treatment, as the heterogeneous nature of the biofilm can complicate the delivery and efficacy of antimicrobial therapies.

Use of antibiotics quickly drives evolution of resistance in both the planktonic and sessile populations, with all planktonic cells developing antibiotic resistance by 4 hours and all sessile cells developing antibiotic resistance between 4 and 8 hours (see Figure (\ref{fig:biofilmtreat})). Use of antibiotics drives formation of new generations as the biofilm is continually destroyed and regrown, shown by the lighter colors of biofilm cells than in the image where no treatment is present. Antibiotics and neutrophils create holes in the biofilm structure. Once all pathogens develop antibiotic resistance, neutrophils are the only agents which can destory the pathogens, and they can be seen eliminating pathogens within the biofilm in the later stages.

On the other hand, Use of phages does not drive as much new division of pathogens or as much mutation of the pathogens as does the use of antibiotics. Note the formation of regions of high phage density on the outer edges of the biofilm at the 12-hour snapshot. This is a result comparable to that found in \cite{Heilmann2012} and can be explained by the fact that along the edges as opposed to deep within the biofilm, phages have a higher adsorption rate and mobility, infected bacteria have a higher capacity to produce phages, and phages have a lower decay rate, an assumption also similar to one made in \cite{Heilmann2012}. Phages also appear more capable than antibiotics of degrading an entire section of biofilm

Overall, a combined use of antibiotics and phages is more effective in quickly degrading a biofilm than either treatment separately. However, if the combined treatment is unable to entirely eliminate the biofilm, it may rebound and continue to develop. Presence of antibiotics still drives evolution of antibiotic resistance as shown in the antibiotic treatment in Figure (\ref{fig:biofilmtreat}), which helps the resurgence of the pathogen population after the first antibiotic dose. After development of antibiotic resistance, only phages and neutrophils can destroy the pathogens. This makes phage treatment an essential counterpart if antibiotic treatment is unsuccessful as shown in the combined treatment in Figure (\ref{fig:biofilmtreat}). Futhermore, in Figure (\ref{fig:biofilmtreat}), phages continue to form high-density regions around the borders of the biofilm, as in \cite{Heilmann2012}

\section{Discussion}
The results of this investigation shed insight on the subtle distinctions between
the outcomes of phage-antibiotic combinations in free-floating and biofilm under
several environmental factors. The protective properties and intricate structure
of biofilms cause major differences in the relative efficacy of therapies. Phage
treatment is useful in reducing pathogen populations in free-floating cultures,
but because of physical barriers and environmental heterogeneity, it is less successful in biofilms. The study’s observation of a large number of phage populations along the biofilm borders supports findings by \cite{Heilmann2012} that showed phages only partially penetrated the biofilm matrix. Compared
to free-floating bacteria, this restriction results in a lower overall impact, which
emphasizes the need for methods to get over biofilm-associated resistance mechanisms.

By developing an agent-based model (ABM) that incorporates intricate microbiological interactions, such as those involving innate immunity, this research
advances the field. The model takes into consideration the intricate structural
and environmental details of biofilms, as well as the interactions between phages,
antibiotics, different bacterial species, and immunological responses. This all-inclusive framework can simulate multiple situations, offering insights into how different treatment approaches and parameter changes affect the proliferation and control of pathogens. The model is a useful tool for investigating the possible efficacy of different therapies and comprehending the interactions between diverse drugs within a biologically realistic setting since it incorporates aspects relevant to human in vivo situations. This work paves the way for future research aimed at optimizing treatment strategies for infections involving biofilms
and highlights the importance of considering a multi-faceted approach in therapeutic development.

In the absence of therapeutic interventions, our observations indicate that
spatial and nutritional constraints limit bacterial growth. This finding aligns
with existing literature suggesting that limited resources can suppress bacterial
proliferation \cite{Monod1949,Gerosa2011}. Our findings demonstrate that the pH range of 3.9
to 8.8 has a uniform impact on the development of biofilms, suggesting that
other factors might be more influential. Nearly all simulations conducted above
19.6 degrees Celsius up to 44.6 degrees Celsius, which can be assumed to be a
natural upper bound for faster enzyme kinetics \cite{Daniel2009,Damico2006}, achieve maximum
population, while the lower end of temperature values from 9.6 degrees Celsius
up to 19.6 degrees Celsius provide faster growth rate. The optimal environment for growth is around 19.6°C. Lower temperatures tend to offer greater stability, but they result in a lower maximum population compared to growth at higher temperatures. We can conclude that bacterial growth in our model is not unlimited, much like in real life.

Our graphs show that when application time is varied, the trends of graphs
for single-phage and single-antibiotic therapy are similar. The highest dosage of
antibiotics forces bacteria to mutate in order to evade them. Therefore, yielding less efficacy compared to lower dosages. Nevertheless, the quantity used
in single-phage therapy strongly correlates with its success rate. The significance of selecting the ideal dosage, particularly in evolutionary trade-offs with
resistance, is reinforced by these two deductions. Crucially, compared to an immunocompromised host, an immunocompetent host facilitates a more efficient
removal of biofilms. This shows that strong therapeutic measures can be less
successful and emphasizes the vital role that a functioning immune response
plays in assisting in the cure of bacterial infections.

Combining the innate immune response with antibiotics and phages proved
to be more effective than treating each condition separately. This discovery
emphasizes the possibility of achieving synergistic benefits through the utilization of multiple therapeutic techniques. The combination therapy appears to
outpace either phage or antibiotic therapy alone, and it also improves the disruption of biofilms. This synergy most likely arises from the complimentary
ways that antibiotics and phages work, which when combined provide a more difficult obstacle for bacterial resistance mechanisms.
Our research also shows that, while to differing degrees of success, all treatment approaches apply pressure to biofilm structures. The innate immune system’s assistance for combined phage-antibiotic therapy is the most efficient way
to compromise the integrity of biofilms. This is explained by the diverse strategy
that combines several methods of disrupting biofilms and killing bacteria. Phage
therapy has a considerable impact on biofilm structure even if it is not as effective as the combined strategy. This is probably because phages are specifically designed to target bacterial cells \cite{fu2010,Kovacs2024}. Conversely, antibiotic therapy is
the least successful in breaking up biofilms; this may be because biofilms have
a high rate of mutation and resistance development \cite{Høiby2010,Mah2010,Stewart2001,Koo2017}. The adaptive ability of biofilms to survive antibiotic stresses is highlighted by observed
resistance, which emphasizes the necessity of combination.

The findings of our study provide a foundation for developing models of microbially induced corrosion inhibition (MICI). While there is extensive research on microbial induced corrosion (MIC) \cite{Enning2014,Videla2005}, MICI remains less understood. Our current model can help determine key tuning parameters and environmental thresholds for effective MICI. Future research will aim to enhance this model by incorporating additional biofilm properties, such as extracellular polymeric substances (EPS), quorum sensing mechanisms, and iron chelation processes, to better understand and optimize biofilm-mediated corrosion inhibition.

In conclusion, our research indicates that although individual treatments
may be successful, the most promising method for controlling biofilms is to
combine phages and antibiotics with support from the innate immune system.
A thorough grasp of biofilm management is provided by the interaction of environmental variables including pH and temperature, the function of host immunity, and the relative efficacy of various treatments. Moreover, clusters, which
have been shown in simulations, might pose new challenges in terms of treatment. In order to better optimize therapy techniques, future study
should investigate the processes underlying the resistance patterns. 

\nolinenumbers


\bibliographystyle{plain}

\newpage
\section*{Appendix}

\section{Implementation of Models}

The agent-based model used for simulations in this paper was developed in the language and programming environment NetLogo (version 6.3.0). Please refer to the commented code provided for more details and to the NetLogo dictionary at \url{https://ccl.northwestern.edu/netlogo/docs/dictionary.html} for additional information on the structure of NetLogo programs.

\subsection{Free-Floating Model}


This version of the program is used for sections 3.3 through 3.7, with some elements switched on or off depending on the case simulated.

The program begins by defining global variables, breeds (agent types), and variables belonging to breeds. The rest of the program consists of a \texttt{setup} procedure, which runs once, followed by a \texttt{go} procedure, which runs continuously.

\subsubsection{Preliminaries}

Some global variables are defined and set in the interface itself so that the user may adjust their values without editing the code. In order to run the code correctly, the user must define these variables on his/her own interface using sliders, switches, inputs, etc. These variables are listed along with their values used in this paper as follows:
\begin{itemize}
    \item \texttt{t}: temperature of environment in degrees Celsius, set to 39.6 for realistic value in human small intestine
    \item \texttt{pH}: pH of environment, set to 6.7 for realistic value in human small intestine
    \item \texttt{mut-chance}: percent chance of a pathogen property mutating upon division, set to 0.25 for value for rate per genome per generation \cite{Drake1998}
    \item \texttt{phage-mut-chance}: percent chance of a phage mutating upon lysis, set to 0.38 for value for rate per genome per generation \cite{Drake1998}
    \item \texttt{burst-size-normal}: base burst size of phages, set to 5 for a scaled-down (computationally viable) value (set to 115 for more biologically realistic value \cite{Drake1998})
    \item \texttt{in-vivo?}: boolean denoting whether simulation represents conditions in human body (i.e., pH automatically regulated around 6.7), set to "on" for realistic human environment
    \item \texttt{A0}: antibiotic dose, set to 50 for a medium value (as in section 3.7)
    \item \texttt{P0}: phage dose, set to 50 for a medium value (as in section 3.7)
    \item \texttt{dose-time}: application time of treatment(s) in ticks, set to 60 (30 minutes) for a medium value (as in section 3.7)
\end{itemize}

The remaining global variables include overall environmental or simulation parameters which are consistent across all agents, such as the ratio $q$ of ticks to minutes (\texttt{q}), the expected phage decay rate per day (\texttt{phage-decay-rate}), and the value around which the body regulates the pH (\texttt{target-pH}); as well as measures kept track of in order to obtain and analyze simulation results, such as the total number of mutations since the start of the simulation (\texttt{num-muts}), the total number of pathogens killed by another type of agent since the start of the simulation (e.g., \texttt{antibiotic-kill-count}), and the average magnitude of the viable temperature range for division across all living pathogens (\texttt{avg-temp-range}). The global variable \texttt{spectrum} defines the spectrum of the antibiotics used in the simulation, where a value of \texttt{0} indicates broad-spectrum and a value of \texttt{1} indicates narrow-spectrum (set in the \texttt{setup} procedure). The provided code contains a full commented list of these variables.

The simulation uses 4 types, or \texttt{breed}s, of agents, named \texttt{pathogen} (pathogenic bacteria, specifically \emph{E. coli}), \texttt{antibiotic} (antibiotic molecules, specifically Rifaximin), \texttt{phage} (bacteriophages, specifically bacteriophage $\lambda$), and \texttt{neutrophil} (representative human immune cells). Agents possess individual variables, which we refer to as properties, whose values may differ among the population. All agents have the properties of health (\texttt{health}), available energy (\texttt{energy}), energy cost of movement (\texttt{move-cost}), and minimum and maximum temperature and pH values for growth (\texttt{tmin, tmax, pHmin, pHmax}).

Pathogens have additional properties for energy cost of survival due to ongoing metabolism (\texttt{survive-cost}), probability of dividing per tick (\texttt{growth-rate-env}, calculated from environmental temperature and pH and individual temperature/pH ranges), generation number (\texttt{gen}), time in ticks since creation (\texttt{generation-time-counter}), antibiotic resistance (\texttt{antibiotic-resistance}, where a value of \texttt{0} indicates susceptible and \texttt{1} indicates resistant), phage resistance (\texttt{phage-resistance}, a list of all phage types to which the pathogen is immune), phage infection status (\texttt{infection}, where a value of \texttt{0} indicates uninfected and \texttt{1} indicates infected), cellular capacity to produce phages (\texttt{kmax}), and ticks until lysis (\texttt{lysis-countdown}). Properties of pathogens that may mutate include \texttt{move-cost}, \texttt{tmin}, \texttt{tmax}, \texttt{pHmin}, \texttt{pHmax}, \texttt{antibiotic-resistance}, and \texttt{kmax}.

Phages have additional properties for burst size (\texttt{burst-size}), latency time of infection (\texttt{latency-time}), probability of dying per tick (\texttt{decay-rate}), adsorption rate (\texttt{adsorption-rate}), and genome type (\texttt{genome-type}, which distinguishes mutated phages). Neutrophils have additional properties for rate of phagocytosis (\texttt{uptake-rate}), lifespan (\texttt{lifespan}), and tick count at creation (\texttt{birthtick}). Antibiotics are characterized by a rate of damage to adjacent pathogens (a number hard-coded into \texttt{kill-bacteria} procedure, not a variable) and a probability of decaying per tick (\texttt{prob-antibiotic-decay}, a global variable). The properties \texttt{health}, \texttt{energy}, \texttt{move-cost}, \texttt{tmin}, \texttt{tmax}, \texttt{pHmin}, and \texttt{pHmax} are irrelevant to all non-pathogen agents and only included for the convenience of using the same procedures for all agents.

\subsubsection{\texttt{setup} Procedure}

The \texttt{setup} procedure clears the previous interface and data, resets the simulation timer (\texttt{ticks}), sets the values of global variables, sets up the environment, and creates the initial population of pathogens. The antibiotic spectrum is specified here, which should be set to \texttt{0} (broad) for Rifaximin. The target pH is set to a realistic value for the human small intestine of 6.7, and the environmental pH is set to that value. Ticks are defined as half-minutes through the parameter $q$. The initial distribution of nutrients is created.

The current environmental and target hydrogen ion concentrations are calculated from the environmental and target pH values using the definition of pH:

\begin{equation} \label{pH-eq}
    [\text{H}^{+}] = {10}^{-pH}
\end{equation}

The individual probabilities of each phage or antibiotic agent decaying are calculated from the expected population-level decay rates using (5) in the Methods and Model Formulation section.

An initial population of $B_0 = 100$ pathogens with random spatial distribution is created and given initial values for properties corresponding to the base values in Table 1. This initial population is defined as generation 0. All of the pathogens begin susceptible to antibiotics and to all phages, and all begin as not infected.

\subsubsection{\texttt{go} Procedure}

The \texttt{go} procedure continuously runs a series of commands and sub-procedures to simulate the growth of bacterial pathogens, the human innate immune response, and treatment with antibiotics and/or phages. The entire procedure runs once every tick after the \texttt{setup} procedure, until either 12 simulated hours (1440 ticks) have passed or all agents have died.

First, the nutrients are replenished randomly across the world through the \texttt{replenish} procedure. Next, an antibiotic dose of $A_0$ is administered if the number of ticks is equal to the application time $t_A$ specified by the global variable \texttt{dose-time} in the interface, or if it is exactly 960 ticks greater than this number (8 hours later, the dose period $\tau_A$). Antibiotic agents are created with a random initial spatial distribution are assigned property values according to Table 1. The environmental hydrogen ion concentration and pH are then updated according to the current abundance of pathogens, which are assumed to increase the hydrogen ion concentration at a rate of ${10}^{-9}$ per pathogen per tick due to natural metabolic processes \cite{Ratzke2018}. The individual probability of each pathogen dividing is set/updated according to the environmental temperature and new pH value and each pathogen's individual properties through (3) in the Methods and Model Formulation section, establishing the new overall pathogen growth rate.

Following this, the main section of the \texttt{go} procedure runs a series of respective sub-procedures on all pathogens, antibiotics, neutrophils, and phages, in that order. An optional lag phase can be added by changing the number of ticks at which these commands begin from 0 to some positive number; until that number of ticks is reached, agents will only move. All results were obtained with no lag phase. After the lag phase, pathogens follow the \texttt{consume}, \texttt{move}, \texttt{try-divide}, \texttt{expire}, and \texttt{lysis} (if infected) procedures in order, and each has its energy reduced by the energy cost of surviving. Antibiotics follow the \texttt{move}, \texttt{kill-bacteria}, and \texttt{dilute} procedures. Neutrophils follow the \texttt{move}, \texttt{kill-bacteria}, and \texttt{expire-neutrophil} procedures. Phages follow the \texttt{move}, \texttt{hijack}, and \texttt{expire-phages} procedures. These procedures are detailed later.

Next, additional global variables are updated which measure total or averaged properties across the simulation. The \texttt{regulate-pH} procedure is used to update the environmental pH if simulating \emph{in vivo} conditions (\texttt{in-vivo?} set to "on" in the interface).

Neutrophils are then created (with random spatial distribution) or removed to meet a population $N_{req}$ commanded by the formula

\begin{equation}
    N_{req} = \lfloor \frac{B - N_{act}}{N_{add}} + 1 \rfloor
\end{equation}

where $B$ is the pathogen population and other variables are as defined in Table 1. If this formula commands a value for $N_{req}$ greater than the maximum allowed neutrophil count $N_{max}$ (due to immune saturation), the commanded value is truncated to $N_{max}$. Neutrophils are assigned property values according to Table 1.

Then, a phage dose of $P_0$ is administered if the number of ticks is equal to the application time $t_A$ specified by the global variable \texttt{dose-time}. Phages are created with a random initial spatial distribution and are assigned property values according to Table 1. All phages are assigned an initial \texttt{genome-type} of 0.

Finally, the number of ticks since the start of the simulation is incremented by 1, and the \texttt{go} procedure repeats.

\subsubsection{Sub-Procedures}

The \texttt{move} procedure is used for all agents, with a separate set of steps for each agent.

\textit{Pathogens:}
To account for the limiting effect of pathogen density on pathogens' movement, pathogens cannot move forward as long as there is another pathogen anywhere in a 90$^\circ$ cone of radius 2 in front of the pathogen. Each pathogen rotates randomly between -60$^\circ$ and 60$^\circ$, and if this cone is free of other pathogens, the pathogen moves forward 1 patch length and expends its energy cost for movement. If this cone is occupied, then the pathogen repeats this process for up to 10 total tries, after which it stays in place and the procedure ends.

\textit{Antibiotics:}
If an antibiotic agent is not blocked by a wall, it rotates randomly between -30$^\circ$ and 30$^\circ$ and moves forward 1 parch length. If it is blocked by a wall, it first rotates to a completely random direction, then rotates right 15$^\circ$ and moves forward 1 patch length if no longer blocked. If still blocked, then the antibiotic agent repeats the 15$^\circ$ rotation and attempts to move forward for up to 10 total tries, after which it stays in place and the procedure ends

\textit{Neutrophils:}
Each neutrophil rotates randomly between -90$^\circ$ and 90$^\circ$ and moves forward 0.5 patch lengths, using an assumption that neutrophils move slower than the other agents modeled.

\textit{Phages:}
Each phage rotates to a completely random direction and moves forward 1 patch length.\\

The \texttt{try-divide} procedure determines whether each pathogen will divide. It carries out the individual probabilities for pathogens dividing, stored in the variable \texttt{growth-rate-env}, by generating a random value between 0 and 1 and allowing the pathogen to potentially divide if the value is less than the pathogen's value of \texttt{growth-rate-env}. A pathogen must also have energy greater than 1, so that pathogens cannot divide forever without gaining energy, as the energy of a pathogen is divided by 2 when it divides. If both of these conditions are satisfied, the pathogen has its energy set to half and its tick counter since creation set to 0. It may then divide via the \texttt{divide} procedure if there are no other pathogens anywhere in a 90$^\circ$ cone of radius 3 in front of the pathogen. If this cone is still occupied, then the pathogen rotates a random degree amount and may then divide if the cone in front of the pathogen becomes unoccupied. It repeats this process for up to 10 total tries, after which it does not divide and the procedure ends.\\

The \texttt{divide} procedure carries out division of pathogens by creating an identical copy of a pathogen after it has passed the \texttt{try-divide} procedure. The new pathogen is created 1 patch length in front of the parent, and it has a separate probability of \texttt{mut-chance} of each of the mutable pathogen properties changing upon division according to the \texttt{mutate-genome} procedures. Aside from these potentially mutated properties, all properties of the parent pathogen are preserved in the offspring. Both parent and offspring become members of the next generation upon division.\\

The \texttt{consume} procedure simulates consumption of nearby nutrients by pathogens. If a pathogen is located on patch containing any sugar, then the energy of the pathogen is incremented by 10, and the sugar count(\texttt{sugar-amount}) of the patch is decreased by 1.

The \texttt{expire} procedure simulates the death of pathogens due to either antibiotic activity, depletion of the pathogens' energy, or pathogens having been alive for a very long time since division. If a pathogen's health or energy falls to or below 0, or if a pathogen has been alive for 500 or more ticks since division, the pathogen dies and is removed from the simulated environment.\\

The set of procedures whose titles begin \texttt{mutate-genome} carry out the mutation of properties of pathogens which may change upon division. Each procedure corresponds to one of the mutable properties listed previously. From the \texttt{divide} procedure, each of these procedures has a separate probability of \texttt{mut-chance} of being called on the new pathogen created in division.

\texttt{mutate-genome1} generates a random integer value from 1 to 10 and then either multiplies or divides the new pathogen's \texttt{movement-cost} value inherited from the parent pathogen by the generated random value, with a probability of 50\% of multiplication or division.

\texttt{mutate-genome2} was originally used to mutate the imposed individual generation time for pathogen division used in early versions of the code. This procedure is empty in the most recent version of the code, but it could be repurposed to simulate mutation of any additional pathogen property.

\texttt{mutate-genome3} generates a random integer value from 1 to 10 and then either increases or decreases the new pathogen's \texttt{tmin} value inherited from the parent pathogen by the generated random value, with a probability of 50\% of addition or subtraction. The same is done for \texttt{tmax}, using a separately generated random value. If \texttt{tmin} is increased beyond the current value of \texttt{tmax}, it is set to the value of \texttt{tmax}. Similarly, if \texttt{tmax} is decreased beyond the current value of \texttt{tmin}, it is set to the value of \texttt{tmin}.

\texttt{mutate-genome4} generates a random value from 0.2 to 2 and then either increases or decreases the new pathogen's \texttt{pHmin} value inherited from the parent pathogen by the generated random value, with a probability of 50\% of addition or subtraction. The same is done for \texttt{pHmax}, using a separately generated random value. (The random values generated are not truly random; instead, they are randomly selected from the set $\{0.2, 0.4, 0.6, 0.8, 1, 1.2, 1.4, 1.6, 1.8, 2\}$.)

\texttt{mutate-genome5} makes the pathogen susceptible to antibiotics if resistant and vice versa; \emph{i.e.}, it switches the value of \texttt{antibiotic-resistance} inherited from the parent pathogen.

\texttt{mutate-genome6} generates a random integer value from 1.1 to 2 and then either multiplies or divides the new pathogen's \texttt{kmax} value inherited from the parent pathogen by the generated random value, with a probability of 50\% of multiplication or division. (The random values generated are not truly random; instead, they are randomly selected from the set $\{1.1, 1.2, 1.3, 1.4, 1.5, 1.6, 1.7, 1.8, 1.9, 2\}$.)\\

The \texttt{kill-bacteria} procedure simulates the depletion of health of pathogens due to antibiotics and the killing of pathogens by neutrophils. Broad-spectrum antibiotics are assumed to be less powerful against pathogens than are narrow-spectrum antibiotics due to their universal use and lack of specific targeting. If the agent following the procedure is a broad-spectrum antibiotic, then if any nearby susceptible pathogens exist, one pathogen is selected and its health is decreased by 25 (1/4 of the maximum pathogen health). If the agent is a narrow-spectrum antibiotic, then if any nearby susceptible pathogens exist, one pathogen is selected and its health is decreased by 50 (1/2 of the maximum pathogen health). If the agent is a neutrophil, then if any nearby pathogens exist, one is selected to die and be removed from the simulated world.\\

The \texttt{dilute} procedure simulates the stochastic dilution and decay of antibiotics by removing the antibiotic agent from the simulation with probability equal to \texttt{prob-antibiotic-decay}, the global probabilistic antibiotic decay rate per tick.\\

The \texttt{replenish} procedure replenishes the simulated world with nutrients every tick by adding 1 to the amount of sugar in 1/8 of the patches which do not already contain the maximum amount of sugar, which is 25. Colors of patches are updated according to the resource distribution using the \texttt{resource-color} procedure.\\

The \texttt{expire-neutrophil} procedure simulates the death of neutrophils. If a neutrophil has been active for a number of ticks equal to or greater than the assigned value of its \texttt{lifespan} property, then it dies and is removed from the world.\\

The \texttt{hijack} procedure simulates the infection of pathogens by phages. A pathogen near the phage is selected, and if such a pathogen exists and is not resistant to the \texttt{genome-type} of the phage, then the pathogen is infected with probability equal to the \texttt{adsorption-rate} of the phage. The pathogen has its \texttt{infection} property set to \texttt{1}, and a countdown until lysis is started from the \texttt{latency-time} of the infecting phage. Upon attempted infection, there is a probability equal to the value of \texttt{prob-acquire-immunity} that the pathogen will become resistant to the \texttt{genome-type} of the infecting phage instead of becoming infected. In this case, the \texttt{genome-type} is added to the list of phage types to which the pathogen is resistant.\\

The \texttt{lysis} procedure simulates the lysis of infected pathogens by phages. It is called on an infected pathogen when the number of ticks since infection is equal to the latency time of the infecting phage. A number of new phages equal to $\lfloor \beta k_{max} \rfloor$ are created at the location of the pathogen, where $\beta$ is the \texttt{burst-size} of the parent phage (or the default burst size if the parent is dead) and $k_{max}$ is the \texttt{kmax} of the pathogen. The pathogen then dies and is removed from the simulated world. New phages are identical to the infecting parent phage, except that each mutates to a brand new \texttt{genome-type} with probability equal to the value of \texttt{phage-mut-chance} divided by 100. The new \texttt{genome-type} assigned is the most recent \texttt{genome-type} number created plus 1.\\

The \texttt{expire-phages} procedure simulates the stochastic decay of phages. Phages die and are removed from the simulated world with probability equal to \texttt{decay-rate}, the probabilistic phage decay rate per tick.\\

The \texttt{regulate-pH} procedure simulates the natural regulation of the pH of the small intestine by the human body. The action of the human body is assumed to approximate proportional control of the hydrogen ion concentration \texttt{p-conc} around the reference value \texttt{target-p-conc}, the hydrogen ion concentration of the small intestine at homeostasis. The hydrogen ion concentration is increased by a value negatively proportional to the error between the target and current hydrogen ion concentrations through the constant \texttt{pH-reg-factor}. The new pH in the simulated world is calculated from the modified hydrogen ion concentration.\\

The \texttt{resource-color} procedure assigns colors to patches in the simulated world based on nutrient distribution. Lighter shades of green are assigned to patches containing less sugar and deeper shades to those containing more. Patches containing no sugar are colored white.\\

The \texttt{mean-gen-times} procedure calculates and displays the mean value of the generation times of all pathogens which have divided since the start of the simulation.\\

\subsection{Biofilm Model}


This version of the program is used for sections 4.4, with some elements switched on or off depending on the case simulated.

Like the free-floating model, the program begins by defining global variables, breeds (agent types), and variables belonging to breeds. The rest of the program consists of a \texttt{setup} procedure, which runs once, followed by a \texttt{go} procedure, which runs continuously.

\subsubsection{Preliminaries}

Some global variables are defined and set in the interface itself so that the user may adjust their values without editing the code. In order to run the code correctly, the user must define these variables on his/her own interface using sliders, switches, inputs, etc. The same user-input global variables as in the free-floating model are used and are set to the same realistic/average values, except that the variable \texttt{max-biofilm} replaces the variable \texttt{dose-time} for determining the first treatment application time. \texttt{max-biofilm} is the number of biofilm cells (sessile cells, pathogens within a biofilm) that must exist before the first treatment dose is applied. It is set to a value of \texttt{0} for application at the beginning of biofilm development, \texttt{300} for application at medium development, and \texttt{1000} for application at full maturity.

All other global variables in the free-floating model are included, and additional global variables are included to implement the variable application time and simulation duration, to report the number of biofilm and infected pathogens, and to establish limits on numbers of agents allowed in the simulation. The variable \texttt{actual-max-biofilm} establishes a maximum of \texttt{1000} biofilm cells to provide a limit on the size of the biofilm at about 1/3 of the simulated space and maintain a free-floating environment in the majority of the simulation. The variable \texttt{max-phages} establishes a maximum of \texttt{35000} phages due to computational constraints, as any number of phages beyond this becomes detrimental to simulation performance. The provided code contains a full commented list of all global variables.

The biofilm model uses the same 4 agent \texttt{breed}s as the free-floating model. All agents have the properties of health (\texttt{health}), available energy (\texttt{energy}), energy cost of movement (\texttt{move-cost}), and minimum and maximum temperature and pH values for growth (\texttt{tmin, tmax, pHmin, pHmax}).

Pathogens have the same additional properties as in the free-floating model, but they are given further properties for probability of adhering to a nearby wall or biofilm cell (\texttt{adhesion-rate}) and for status of being within a biofilm (\texttt{biofilm?}, where a value of \texttt{0} indicates free-floating and \texttt{1} indicates biofilm). Like the free-floating model, properties of pathogens that may mutate include \texttt{move-cost}, \texttt{tmin}, \texttt{tmax}, \texttt{pHmin}, \texttt{pHmax}, \texttt{antibiotic-resistance}, and \texttt{kmax}.

Phages have the same additional properties as in the free-floating model, but they are given further properties for probability of converting a nearby biofilm cell to a free-floating pathogen (\texttt{EPS-degradation-rate}, representing degradation of the EPS matrix) and for the distance in patch lengths (2.0 $\mu$ m) which a phage may travel each tick (\texttt{phage-mobility}, affected by proximity to biofilm cells). Neutrophils and antibiotics have the same additional properties as in the free-floating model. The properties \texttt{health}, \texttt{energy}, \texttt{move-cost}, \texttt{tmin}, \texttt{tmax}, \texttt{pHmin}, and \texttt{pHmax} are irrelevant to all non-pathogen agents and only included for the convenience of using the same procedures for all agents.

\subsubsection{\texttt{setup} Procedure}

The \texttt{setup} procedure clears the previous interface and data, resets the simulation timer (\texttt{ticks}), sets the values of global variables, sets up the environment, and creates the initial population of pathogens. The \texttt{setup} procedure is identical to that in the free-floating model, aside from setting the values of the additional global variables described above. All pathogens additionally begin as planktonic (free-floating).

\subsubsection{\texttt{go} Procedure}

The \texttt{go} procedure continuously runs a series of commands and sub-procedures to simulate the growth of bacterial pathogens, the development of a biofilm, the human innate immune response, and treatment with antibiotics and/or phages. The entire procedure runs once every tick after the \texttt{setup} procedure, until either 12 simulated hours (1440 ticks) have passed since the first dose or all agents have died. The \texttt{go} procedure is similar to that in the free-floating model but has differences regarding the variable dose time, distinct behavior of free-floating and biofilm pathogens, and behavior of phages.

First, the nutrients are replenished randomly across the world through the \texttt{replenish} procedure. Next, an antibiotic dose of $A_0$ is administered if the number of biofilm cells is greater than or equal to the number chosen for the first dose (\texttt{max-biofilm}) and no doses have been applied. An antibiotic dose of $A_0$ is also administered if exactly 960 ticks (8 hours, the dose period $\tau_A$) have passed since the time of the first dose. Antibiotic agents are created with a random initial spatial distribution are assigned property values according to Table 1. The environmental hydrogen ion concentration and pH are then updated according to the current abundance of pathogens, which are assumed to increase the hydrogen ion concentration at a rate of ${10}^{-9}$ per pathogen per tick due to natural metabolic processes \cite{Ratzke2018}. The individual probability of each pathogen dividing is set/updated according to the environmental temperature and new pH value and each pathogen's individual properties through (3) in the Methods and Model Formulation section, establishing the new overall pathogen growth rate.

Following this, the main section of the \texttt{go} procedure runs a series of respective sub-procedures on all pathogens, antibiotics, neutrophils, and phages, in that order. An optional lag phase can be added by changing the number of ticks at which these commands begin from 0 to some positive number; until that number of ticks is reached, agents will only move. All results were obtained with no lag phase. After the lag phase, pathogens follow the \texttt{consume}, \texttt{move}, \texttt{unadhere}, \texttt{try-divide} (if planktonic), \texttt{expire}, and \texttt{lysis} (if infected) procedures in order, and each has its energy reduced by the energy cost of surviving, where a cost of 0.25 is used for biofilm cells and a cost of 0.5 is used for planktonic cells. Antibiotics follow the \texttt{move}, \texttt{kill-bacteria}, and \texttt{dilute} procedures. Neutrophils follow the \texttt{move}, \texttt{kill-bacteria}, and \texttt{expire-neutrophil} procedures. Phages follow the \texttt{adjust-properties}, \texttt{move}, \texttt{hijack}, \texttt{degrade-EPS}, and \texttt{expire-phages} procedures. Next, if the biofilm is not yet at maximum capacity, then planktonic pathogens follow the \texttt{adhere} procedure (attempt to adhere to wall or biofilm cell) and biofilm pathogens follow the \texttt{try-divide} procedure (attempt to divide). All of these procedures are detailed later.

Next, additional global variables are updated which measure total or averaged properties across the simulation. The \texttt{regulate-pH} procedure is used to update the environmental pH if simulating \emph{in vivo} conditions (\texttt{in-vivo?} set to "on" in the interface).

Neutrophils are then created in a process identical to that in the free-floating model.

Then, a phage dose of $P_0$ is administered only if the first antibiotic dose was administered the same tick (thus, the first antibiotic dose and the phage dose are always applied simultaneously). Phages are created with a random initial spatial distribution and are assigned property values according to Table 1. All phages are assigned an initial \texttt{genome-type} of 0 and have default mobility.

The procedure \texttt{dilute-planktonic} then removes a number of the planktonic pathogens proportional to their population, to prevent crowding of the simulation and model a constant bacterial density in the environment surrounding the developing biofilm.

Finally, the number of ticks since the start of the simulation is incremented by 1, and the \texttt{go} procedure repeats.

\subsubsection{Sub-Procedures}

The \texttt{move}, \texttt{try-divide}, \texttt{divide}, \texttt{consume}, \texttt{expire}, \texttt{mutate-genome} (1 through 6), \texttt{kill-bacteria}, \texttt{dilute}, \texttt{replenish}, \texttt{expire-neutrophil}, \texttt{hijack}, \texttt{lysis}, \texttt{expire-phages}, \texttt{regulate-pH}, \texttt{resource-color}, and \texttt{mean-gen-times} procedures are identical to those described for the free-floating model, aside from the differences listed below:\\

\begin{itemize}
\item The \texttt{move} procedure for pathogens only proceeds if the pathogen is planktonic (not in a biofilm).

\item The \texttt{try-divide} procedure applies separately to planktonic and biofilm cells.

\begin{itemize}
    \item \textit{Planktonic:} If a pathogen is planktonic and all other conditions (described for the free-floating model) are satisfied, the pathogen may then divide via the \texttt{divide} procedure if there are no other pathogens anywhere in a 90$^\circ$ cone of radius 2 in front of the pathogen. If the cone is still occupied, then the pathogen rotates a random degree amount and may then divide if the cone in front of the pathogen becomes unoccupied. It repeats this process for up to 10 total tries, after which it does not divide and the procedure ends.
    \item \textit{Biofilm:} If a pathogen is in a biofilm and all other conditions (described for the free-floating model) are satisfied, the pathogen may then divide via the \texttt{divide} procedure if there are no other pathogens anywhere in a 60$^\circ$ cone of radius 2 in front of the pathogen. If the cone is still occupied, then the pathogen rotates right 15$^\circ$ and may then divide if the cone in front of the pathogen becomes unoccupied. It repeats this process for up to 10 total tries, after which it does not divide and the procedure ends.
\end{itemize}

\end{itemize}

The following additional sub-procedures are included in the biofilm model:\\

The \texttt{adhere} procedure simulates the adhesion of planktonic pathogens to the bottom wall of the simulated world or to pathogens in a biofilm. If a planktonic pathogen is within 1 patch length of the bottom wall or of at least one biofilm cell, then it becomes a biofilm cell with probability equal to the \texttt{adhesion-rate}. If it becomes a biofilm cell, its color is changed to purple and its phage production capacity \texttt{kmax} is reduced by half to model the negative effect of dormancy in biofilm cells on phage replication.\\

The \texttt{unadhere} procedure simulates the detachment of biofilm cells from a biofilm, in the case that the cells have become separated from the biofilm through pathogen death or biofilm degradation. If a biofilm cell is no longer surrounded by any biofilm cells, then it becomes a planktonic pathogen, has its color changed to red, and has its phage production capacity \texttt{kmax} doubled to model the positive effect of a planktonic status on phage replication.\\

The \texttt{dilute-planktonic} procedure carries out the deterministic proportional dilution of the planktonic pathogens, which is imposed to prevent crowding of the simulation and model a constant bacterial density in the environment surrounding the developing biofilm. Each tick, 1.7\% of planktonic pathogens are randomly selected to die and be removed from the simulated world.\\

The \texttt{adjust-properties} procedure adjusts the properties of phages depending on their presence within a biofilm, which is assessed through the number of nearby biofilm cells \cite{Heilmann2012}. If any biofilm cells are within a radius of 1 unit of the phage, then the \texttt{decay-rate} of the phage is increased by 20\% for each biofilm cell within a radius of 1, and the \texttt{adsorption-rate} and \texttt{phage-mobility} of the phage are each decreased by 20\% for each biofilm cell within a radius of 1. If there are no pathogen cells within a radius of 1 of the phage, then the \texttt{decay-rate}, \texttt{adsorption-rate}, and \texttt{phage-mobility} of the phage are set to their normal values, as given for the phages initially created in the simulation.\\

The \texttt{degrade-EPS} procedure simulates the degradation of the EPS matrix and thus of biofilms by phages near the boundary of a biofilm. If either 1 or 2 biofilm cells is within a radius of 1 unit of the phage, then one of the biofilm cells is converted to a planktonic pathogen with probability equal to the value of the \texttt{EPS-degradation-rate} of the phage. The converted pathogen has its color changed to red and has its phage production capacity \texttt{kmax} doubled to model the positive effect of a planktonic status on phage replication.\\

\section{Formula Derivations}

\subsection{Bacterial Relative Growth Rate Formula}

The square-root type model developed in \cite{Ross2003} for the growth of \emph{E. coli} as a function of temperature $T$, water activity $a_w$, pH $pH$, and lactic acid concentration $LAC$ was adopted in order to describe the effect of environmental factors on growth rate. Temperature and pH were selected as significant factors in bacterial growth for our model, so the original function was simplified to a function of only temperature and pH by excluding terms related to water activity and lactic acid concentration. This resulted in the following growth rate function:

\begin{equation} \label{sup1}
    f(T, pH) = {\left( 0.043885 (T - T_{min}) \left( 1 - \exp{ \left( 0.2636 (T - T_{max}) \right) } \right) \right) }^{2} \left(1 - {10}^{pH_{min} - pH}\right) \left(1 - {10}^{pH - pH_{max}}\right)
\end{equation}

Since this is a function for relative growth rate, we can derive the following relationship with a function $g(T, pH)$ representing the equivalent proportion of the population that divides each tick as follows.

\begin{itemize}
    \item let $t$ represent time in hours
    \item let $\tau$ represent time in ticks
    \item let $q$ represent $\tau$/$t$
    \item let $p$ represent the colony's population
    \item let $r_t$ represent relative growth rate in 1/hours, defined by \[r_t = \frac{1}{p} \cdot \frac{dp}{dt}\]
    \item let $f(\Phi)$ represent the relative growth rate $r_t$ as a function of environmental parameters $\Phi=\{T,pH,...\}$, as determined experimentally in Ross et al, 2003
    \item let $g(\Phi)$ represent the probability of a cell dividing each tick, which is equivalent to the expected proportion of the population that divides
\end{itemize}

\[
p = p_0(1 + g(\Phi))^\tau
\]

\[
\begin{split}
    \frac{dp}{d\tau} &= p_0(1 + g(\Phi))^\tau \cdot \ln (1+g(\Phi))\\
    &= p \cdot \ln (1+g(\Phi))
\end{split}
\]
    
\[
\begin{split}
    \ln (1+g(\Phi)) &= \frac{1}{p} \cdot \frac{dp}{d\tau} \\
    &= \frac{1}{p} \cdot \frac{dp}{dt} \cdot \frac{dt}{d\tau} \\
    &= {r_t} \cdot \frac{1}{q}  \\
    &=  \frac{1}{q} \cdot f(\Phi)
\end{split}
\]

\[
g(\Phi) = e^{\frac{1}{q} \cdot f(\Phi)} - 1
\]

\begin{itemize}
    \item For $g(\Phi)$ to make sense as a probability, we require that $ f(\Phi) \le q \ln 2 $ for all $\Phi$ so that $ g(\Phi) \le 1 $
\end{itemize}

NOTE: Relationship holds in general for any deterministic and probabilistic growth rates in units of time with ratio of duration given by $q$.
\\\\
For a rate of decay ($f$ is relative decay rate), the resulting relationship is:

\[
g(\Phi) = 1 - e^{- \frac{1}{q} \cdot f(\Phi)}
\]

These results are supported by the ABM model formulation described in the supplement for Heilmann et al, 2012.

\subsection{Timescale Restriction}

We have the following formula for $g$:

\[
g(\Phi) = e^{\frac{1}{q} \cdot f(\Phi)} - 1
\]

\begin{itemize}
    \item Therefore, for $g(\Phi)$ to make sense as a probability, we require that $ f(\Phi) \le q \ln 2 $ for all $\Phi$ so that $ g(\Phi) \le 1 $
\end{itemize}

\subsection{Decay Formula}

To convert an hourly decay rate to an individual probability of decaying per tick, we used the formula derived below.

\begin{itemize}
    \item let $t$ represent time in hours
    \item let $\tau$ represent time in ticks
    \item let $q$ represent $\tau$/$t$
    \item let $p$ represent the total population
    \item let $\delta_t$ represent the decay rate in 1/hours
    \item let $\delta_\tau$ represent the decay rate in 1/ticks, equivalent to the probability of an individual agent decaying each tick, or to the expected proportion of the population which decays each tick
    \item Note: the decay rates above are NOT relative decay rates, i.e., not equivalent to $-\frac{1}{p} \cdot \frac{dp}{dt}$
\end{itemize}

We have

\[
p = p_0 {(1 - \delta_t)}^t
\]

and

\[
p = p_0 {(1 - \delta_\tau)}^\tau
\]

Setting these equal and substituting $\tau = qt$,

\[
{(1 - \delta_t)}^t = {(1 - \delta_\tau)}^{qt}
\]

\[
1 - \delta_\tau = {(1 - \delta_t)}^{1/q}
\]

\[
\delta_\tau = 1 - {(1 - \delta_t)}^{1/q}
\]

Therefore, in order to match an hourly decay rate of $\delta_t$, the probability of an agent decaying/dying each tick should be set as

\[
\delta_\tau = 1 - {(1 - \delta_t)}^{1/q}
\]

\end{document}